\documentclass[onecolumn,authoryear]{els-mrw} 

\usepackage{amsmath,amssymb,amsfonts,amsthm,makeidx,graphicx}
\usepackage{txfonts}
\usepackage{helvet}
\usepackage{aas_macro}
\usepackage{siunitx}

\DeclareSIUnit \solarmass {\ensuremath{M_\odot}}

\newcommand{\mbh}{M_{\rm BH}}
\newcommand{\mstar}{M_{*}}
\newcommand{\ssigma}{\sigma_{*}}
\newcommand{\mbulge}{M_{\rm bulge}}
\newcommand{\fedd}{\lambda_\mathrm{Edd}}
\newcommand{\Msun}{M_\odot}

\def\oiii		{$\mathrm{\left[ O \textsc{iii}\right] }$}

\def\halpha            {{\rm H}\alpha}

\begin{document}

\chapter{Massive black holes and their galaxies}\label{chap1}

\author[1]{Ricarda S. Beckmann}%
\author[2]{Rebecca J. Smethurst}%

\address[1]{\orgname{University of Edinburgh}, \orgdiv{Royal Observatory Edinburgh, Institute for Astronomy}, \orgaddress{Blackford hill, Edinburgh EH9 3HJ, UK}}
\address[2]{\orgname{University of Oxford}, \orgdiv{Oxford Astrophysics, Department of Physics}, \orgaddress{Denys Wilkinson Building, Keble Road, Oxford, OX1 3RH, UK}}

\articletag{Author accepted manuscript of "Massive black holes and their galaxies" from the Encyclopedia of Astrophysics, 1st Edition.}

\maketitle

\begin{glossary}[Glossary]
\term{Massive black hole (MBH)} A black hole with a mass of $10^3 \rm \ \Msun$ or more \\
\term{Intermediate-mass black hole (IMBH)} A black hole with a mass of $10^3 \rm \ \Msun < \mbh < 10^6 \rm \ \Msun$ \\
\term{Supermassive black hole (SMBH)} A black hole with a mass of $\mbh > 10^ 6 \rm \ \Msun$ \\
\term{Active galactic nucleus (AGN)} An actively accreting MBH observable in electromagnetic radiation \\
\term{Seed black hole} A newly formed black hole which grew into the MBH we observe \\
\term{Quasar} A radiatively efficient AGN \\
\term{``quasar mode ''} MBH accretion through a radiatively efficient accretion disc  \\
\term{``jet mode''} MBH accretion through a radiatively inefficient accretion disc that drives a jet \\
\term{bulge-dominated galaxy} A spheroidal, dispersion dominated galaxy. Also called an early-type or ellptical galaxy \\
\term{disc-dominated galaxy} A rotation dominated galaxy. Also called a late-type galaxy \\

\end{glossary}

\begin{glossary}[Nomenclature]
%\rbc{Struggling a bit with the nomenclature here. What do we go for for galaxy types? Early-type versus late-type? I have stuck with that for now as much as I hate that classification.}\rjs{I also hate that nomenclature - disk-dominated and bulge-dominated could be an alternative?}\rjs{24-11-24: I have gone through and had a re-arrange to put content that was there into a more coherent order. I have added bits and bobs as I went. I've made notes where I still want to add more. Also I desperately want to hyphenate all instances of coevolution.} \rbc{I am fairly sure it is coevolution! Why would you hyphenate it? }
\begin{tabular}{@{}lp{34pc}@{}}
$\mbh$ & BH mass \\
$\mstar$ & Stellar mass of a galaxy \\
$\mbulge$ & Stellar mass of the bulge of a galaxy \\
$\ssigma$ & Stellar velocity dispersion of a galaxy \\
$L_{\rm AGN}$ & The bolometric luminosity of an AGN. Typically measured in $\rm erg \ s^{-1}$. \\
$\fedd$ & The Eddington ratio of an actively growing BH. Higher Eddington ratios denote more efficient accretion. \\
\end{tabular}
\end{glossary}

\begin{abstract}[Abstract]
Almost every galaxy in the local Universe is observed to have a massive black hole in the centre. The properties of these black holes are observed to tightly correlate with those of their host galaxy which has been interpreted as coevolution regulated by black hole feedback. This coevolution spans most of cosmic history, as the first active black holes, so-called active galactic nuclei, are already observed as early as $z\sim10$. In this chapter, we lay out how we can find supermassive black holes, review what we know about the population of black holes and their host galaxies from observations, and summarise what we have learned about their coevolution across cosmic time from both observations and simulations. 
\end{abstract}

%\section*{Chapter outline}
%\extit{To be assembled in typesetting}

\section*{Objectives}
In this chapter we will

\begin{itemize}
    \item Recount how we first discovered massive black holes and their link to their host galaxy
    \item Outline how we observe massive black holes and measure their masses
    \item Review the properties of massive black holes in the local Universe
    \item Present the observational evidence for the coevolution of massive black holes and galaxies
    \item Investigate how massive black hole growth has evolved over cosmic time
    \item Summarise what we know about galaxy-scale processes that fuel massive black holes
    \item Present how feedback energy from massive black holes shapes galaxy evolution
\end{itemize}

\vspace{-0.2cm}
\section{Introduction}
Almost every massive galaxy in the local Universe is observed to have a massive black hole (MBH) in the centre. A variety of evidence now suggests that MBH and their host galaxies not only co-exist but co-evolve: the environment of the galaxy shapes the growth of the black hole, and the feedback energy from the growing BH shapes the evolution of the host galaxy. Coevolution spans most of cosmic history, as the first active BH, so called active galactic nuclei (AGN), are already observed as early as $z\sim10$ \citep{bogdan2024EvidenceHeavyseedOrigin, Whalen2024UHZ1}.

\section{A  brief history of the discovery of massive black holes and their coevolution with galaxies}\label{sec:history}

 %In this chapter, we review what we know about the population of MBH and summarise what we have learned about their coevolution across cosmic time from both observations and simulations. 

The first AGN was observed by \citet{fath1909SpectraSpiralNebulae}, who noted emission and absorption lines reminiscent of gaseous nebulae rather than stellar emission. By the time \citet{seyfert1943NuclearEmissionSpiral} was studying galaxies, he suggested the observed line widths were due to Doppler broadening from high ($>10^3 \rm \ km/s$) rotational velocities, which implied a massive central object. The detection of the radio background by \citet{jansky1932DirectionalStudiesAtmospherics} launched a series of active radio observations in the 1950s that uncovered both extended and compact radio sources and gave rise to the idea that they might be extragalactic  \citep{baade1954IdentificationRadioSources,shields1999BriefHistoryActive}. 
Soon after, quasi-stellar objects (``quasars") were recognised as extremely bright, extragalactic objects following the unexpectedly high redshift of what we now know as the first quasar, 3C 273 \citep[$z\sim0.158$;][]{schmidt19633C273StarObject}. From a consideration of the energy required to power such an object, it was suggested that 3C 273 was powered by an extremely dense, massive central object of $\sim 10^9  \rm \ M_\odot$ \citep{greenstein1964QuasiStellarRadioSources}. It was \citet{lynden-bell1969GalacticNucleiCollapsed} who then identified the origin of the energy to be gravitational, which gave rise to the argument from \citet{sotan1982MassesQuasars} that the high space density of quasars meant ``good prospects for the observational detection of the quasar remnants in nearby galaxies", prompting a search for such objects and a characterisation of their masses. 

The most direct way to measure the mass of a MBH is to use the Keplerian rotation of baryonic matter in their gravitational potential. The first such dynamical mass measurement, using stars at the centre of M32, was reported by \citet{tonry1984EvidenceCentralMass}. According to \citet{kormendy2013CoevolutionNotSupermassive} this was still a "marginally possible measurement" where the errorbars on the data points were so large that it was only just feasible to detect the BH at all. Around the same time, theoretical efforts focused on developing a coherent understanding of how such massive BH might form and evolve. The model presented by \citet{rees1984BlackHoleModels} already contained many of the core features of our current model of MBH evolution.

By the 1990s, the properties of a wide range of different AGN had been studied in detail \citep{osterbrock1991ActiveGalacticNuclei} and the MBH model of AGN was well established, if still under discussion in the details \citep{blandford1992StandardModelNew}. Improved sensitivity of ground based optical observatories in the 1990s improved mass measurements, culminating with the launch of the Hubble Space Telescope which resolved the host galaxies of distant quasars. As soon as a large enough sample of MBH mass measurements had been assembled it was then noted that MBH masses appear to be tightly linked to different properties of their host galaxies. The first correlation between the stellar mass of the galaxy bulge and the MBH mass was reported by \citet{magorrian1998DemographyMassiveDark}, but a small sample and large scatter made drawing strong conclusions difficult (although note the MBH was referred to as ``massive dark object", or MDO, throughout \citealt{magorrian1998DemographyMassiveDark}, and that the identification of their MDOs as BH or otherwise was ``not important for the
purposes of this paper"). In 2000, \citet{ferrarese2000FundamentalRelationSupermassive} and \citet{gebhardt2000RelationshipNuclearBlack} simultaneously reported a tight correlation between the central MBH mass and the stellar velocity dispersion,  $\sigma$ of the central bulge. Soon after, updated mass measurements showed that the correlation between MBH mass and host galaxy stellar bulge mass was also tighter than originally reported \citep{haring2004BlackHoleMassBulge}. Since then, less tight correlations have been reported between the mass of the MBH and total galaxy stellar mass \citep{reines2015RELATIONSCENTRALBLACK}, spiral arm pitch angle \citep{Seigar2008, Berrier2013, Davis2017}, and a `fundamental plane' \citep{Wong2016, Wang2024}. The tightness of these correlations, known in the literature as \emph{scaling relations}, is one of the key arguments for the coevolution of galaxies and their central MBH: the galaxy shapes the BH and the BH shapes the galaxy.

While for many the argument for the existence of central MBH in galaxies was becoming increasingly convincing, the evidence remained circumstantial. It relied on concluding that the inferred mass density could only be a MBH, as opposed to a swarm of stellar BH or other massive dark objects. Tightening constraints came from measuring the mass of the Milky Way's central black hole, SgrA*, using direct imaging of stellar orbits \citep{ghez2008MeasuringDistanceProperties,genzel2010GalacticCenterMassive}. Over a decade later, direct imaging of the accretion discs around the MBH in M87 and SgrA* confirm the theory that AGN are powered by the accretion discs of MBH \citep{event2019FirstM87Event}.

Note that traditionally, BH with masses of $M_{\rm BH} > 10^6 \rm \ M_\odot$ are referred to as supermassive BH (SMBH), while those in the mass-range $ 10^3 M_\odot < M_{\rm BH} < 10^6 \rm \ M_\odot$ are referred to as intermediate mass BH (IMBH). Anything below is considered a stellar mass BH. In this chapter we use the term ``massive black hole" (MBH) to refer to all central galactic BH with a mass of more than $10^3 \rm \ M_\odot$ and use the terms IMBH and SMBH only when specifically applicable to that population.

\section{Finding and characterizing massive black holes}
\label{sec:BHmasses}

One of the main difficulties in studying the coevolution of MBH and their host galaxies is the fact that MBH are only detectable when they are either actively growing, as AGN, or when they are sufficiently nearby for us to be able to measure the imprint of their gravitational potential on nearby gas and stars. This inherently restricts sample sizes. In this section we briefly discuss both direct and indirect methods of measuring MBH masses and how their limitations shape our question for understanding coevolution.

\subsection{Direct dynamical supermassive black hole mass measurements}
\label{sec:dynamical_mass}
The most direct method to detect MBH, and measure their masses, is to look for the imprint their gravitational potential on the orbits of nearby baryonic matter. As this method requires measuring gas or stellar dynamics quite close to the black hole, this method is restricted to nearby galaxies with quiescent AGN. Following the method first described in \citet{schwarzschild1979NumericalModelTriaxial}, the BH mass is measured by numerically integrating the equation of motion for a representative library of orbits, which have to take the potential of the galaxy, the dark matter halo and the central MBH into account. The BH mass is constrained by finding the set of orbits which best fit the combined density and velocity moments of the observed surface brightness and kinematics of the host galaxy. 

Except for Sagittarius A*, where direct imaging of stellar orbits as close as 500 Schwarzschild radii is possible \citep{ghez2008MeasuringDistanceProperties, genzel2010GalacticCenterMassive}, even highly accurate stellar dynamical mass measurements typically probe only as close as $10^4 - 10^5$ Schwarzschild radii from the BH. Uncertainty is introduced in the estimate from the unknown anisotropy in the distribution of central stars, the tri-axiality of the galaxy's potential, variations in the mass-to-light ratio, and difficulties in robustly identifying the location of the potential BH \citep{greene2020IntermediateMassBlackHoles}.

A similar analysis is possible by studying ionised nebular emission coming from a central gas disc around the MBH that is assumed to be Keplerian. As well as sharing many of the sources of uncertainty of stellar dynamical mass measurements, observations of ionised gas near MBH show unexpectedly large line widths, the interpretation of which is unclear \citep{kormendy2013SecularEvolutionDisk}. For this reason, ionised-gas measurements are considered less reliable than stellar-dynamical ones. Finally, stimulated spectral line emission from ``masers'' can be used to probe rotation in galaxies and constrain BH masses. While this method has been shown to give robust mass measurements, \citep{miyoshi1995EvidenceBlackHole,kuo2011MEGAMASERCOSMOLOGYPROJECT}, the overall sample size of BH masses measured using this method remains small \citep{farhan2023WaterMegamaserCentral}. Dynamical mass measurements preferentially detect quiescent MBH, so those not currently actively growing.

\subsection{Indirect massive black hole masses for active galactic nuclei}
\label{sec:AGNmasses}
Dynamical mass measurements are only possibly in nearby galaxies. To understand the wider population of MBH, and their evolution over cosmic time, we need to estimate the MBH masses and growth rates for large samples of AGN across different redshifts.  A range of methods have been developed to indirectly measure BH masses from AGN spectra; the two most widely used are reverberation mapping and single-epoch mass estimators.

Reverberation mapping relies on inferring the rotational velocity of the accretion disc which, as with dynamical mass measurements, can be used to measure the BH mass  \citep{peterson2004CentralMassesBroadLine}. This method uses the fact that the strength of the broad lines, emitted from the outer parts of the MBH accretion disc, depend on the flux from the central source. By measuring the time-delay between variations in the continuum (from near the MBH) and the broad lines, one can infer the distance from the MBH to the broad line region \citep{blandford1982ReverberationMappingEmission}. This method involves a dimensionless scaling parameter, which has to be calibrated from dynamical mass measurements. While reverberation mapping is thought to be very accurate, repeated spectra at long intervals of the same source are required \citep{peterson2013MeasuringMassesSupermassive}, which is observationally expensive \citep{mclure2002MeasuringBlackHole}. Unfortunately, the variable transmissivity of Earth's atmosphere for optical lines makes such observations a challenge from the ground \citep{peterson1993ReverberationMappingActive}. Other sources of uncertainty include the assumption that any changes in flux are isotropic and the fact that each emission line has a different transfer function for how it responds to a delta-function burst of radiation. 

Fortunately, reverberation mapped MBH revealed that there is a tight empirical relation between the size of the broad-line region and the optical luminosity of the AGN. By calibrating from this relation, it is possible to estimate the BH mass by measuring the optical luminosity and the width of the broad lines, e.g. EW[$\halpha$], from a single AGN spectrum. This economical and efficient method is called the single-epoch mass estimator \citep{greene2005, kaspi2005RelationshipLuminosityBroadLine, bentz2013LOWLUMINOSITYENDRADIUSLUMINOSITY} and it allows for MBH mass estimates across a wide range of redshifts, environments, and galaxy masses. Large spectroscopic galaxy surveys have resulted in single-epoch mass measurements becoming the most commonly used estimator of MBH mass across the literature. 

The main source of uncertainty of single-epoch mass estimators is that it relies on reverberation-mapping calibrated measurements, which in turn rely on dynamical mass measurements. Errors and uncertainties are propagated through these dependencies. While direct methods of MBH measurements produce estimates with uncertainties within 0.5 dex \citep{williams2023AssessingIndirectMethods,peterson2013MeasuringMassesSupermassive}, indirect methods, such as single-epoch mass estimators, produce estimates with uncertainties up to 1 dex  \citep{gliozzi2024ComparingIndirectMethods}. 

Other methods for indirect MBH mass measurements include X-ray variability \citep{nikolajuk2004BlackHoleMass,akylas2022BlackHoleMass,zhou2010CALIBRATINGCORRELATIONBLACK} and the X-ray spectral index \citep{gliozzi2011TESTINGSCALEINDEPENDENTMETHOD}, along with the fundamental plane of BH activity which empirically links X-ray and radio emission to MBH mass \citep{merloni2003FundamentalPlaneBlack,falcke2004SchemeUnifyLowpower}. In principle one can also detect BH through microlensing events, where background light is deflected by the BH's gravitational potential \citep{lam2022IsolatedMassgapBlack}, but so far microlensing searches have only detected stellar mass BH \citep{wyrzykowski2020ConstrainingMassesMicrolensing}. Similarly direct detection of gravitational waves (emitted when two BH or neutron stars merge) with interferometers such as VIRGO and LIGO, have detected stellar mass BH \citep{abbott2023GWTC3CompactBinary}. Pulsar timing arrays have indirectly detected a diffuse gravitational wave background which could be explained through a large number of blended MBH mergers \citep{antoniadis2022InternationalPulsarTiming, EPTA2023}. 

Finally, tidal disruption events have been used to detect MBH and measure their masses: mass estimates are obtained by comparing observed light-curves to those predicted by hydrodynamical simulations with different stellar and BH parameters \citep{mockler2019WeighingBlackHoles}. Stars are only disrupted in tidal disruption events below a maximum MBH mass, the so-called Hills mass \citep{hills1975PossiblePowerSource}, which depends on the disrupted star's density and the spin of the MBH. At higher MBH masses stars are accreted whole, which does not produce the characteristic light curve of a tidal disruption event. For a typical main-sequence star the Hills mass is of order $10^8 \rm \Msun$ \citep{hills1975PossiblePowerSource,kesden2012TidaldisruptionRateStars}. For much more compact white dwarfs, it can be as low as $\sim 2 \times 10^5 \rm \Msun$ \citep{gezari2021TidalDisruptionEvents}. Measuring MBH masses via tidal disruption events therefore offers exciting future possibilities to extend the MBH sample to lower masses, and to probe the population of otherwise quiescent SMBH and IMBH.

Finally, it is in principle possible to estimate MBH masses from galaxy properties using observed empirical scaling relations (see Sec.~\ref{sec:history} and~\ref{sec:scaling_relations}). However, there is a large scatter around many of these scaling relations which results in inaccurate MBH mass estimates with large uncertainties in comparison to other methods \citep{gliozzi2024ComparingIndirectMethods}. As such, we do not recommend using scaling relations to obtain MBH masses. 

\subsection{Massive black hole masses are tightly correlated with their host galaxy properties}
\label{sec:scaling_relations}

\begin{figure}
    \centering
    \includegraphics[width=\linewidth]{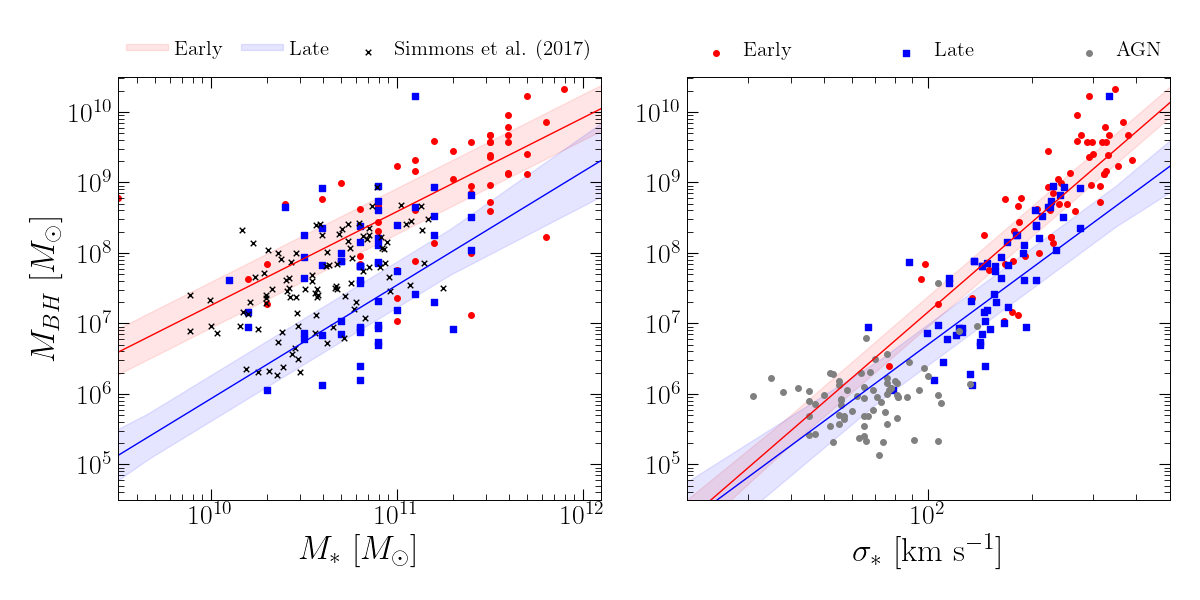}
    \caption{Scaling relations for MBH and galaxies in the local Universe, showing the $\mbh-\mstar$ relation (left) and the $\mbh-\ssigma$ relation (right). Red denotes bulge-dominated galaxies while blue is for disc-dominated galaxies. Black crosses denote the bulge-less galaxies from \citep{simmons2013GalaxyZooBulgeless}. Grey and black markers are for AGN while coloured MBH masses are measured using direct detection methods. Courtesy of J. Greene, using data published in published in \citep{greene2020IntermediateMassBlackHoles}.}
    \label{fig:scaling_relations}
\end{figure}

With measured masses of MBH secured, one can then study their relationship with their host galaxies. As can be seen in Fig.~\ref{fig:scaling_relations}, more massive galaxies, with higher stellar velocity dispersions, host more massive BH. 
These scaling relations are one of the key pieces of evidence that MBH and galaxies do not just coexist but coevolve. The tightness of these relations goes far beyond what we would expect from pure hierarchical structure formation. This is especially true as MBH masses make up only about 0.1 \% of the total stellar mass of the galaxy.

Given the limitations to MBH mass measurements discussed above, a key caveat to consider before drawing conclusions from such observations like those shown in Fig.~\ref{fig:scaling_relations} is how complete the plotted sample is. All data-points not labelled as AGN in Fig.~\ref{fig:scaling_relations} are from dynamical mass measurements (see Sec.~\ref{sec:dynamical_mass}) and therefore necessarily from local galaxies alone. \citet{shankar2016SelectionBiasDynamicallymeasured} showed that galaxies for which direct mass measurements are possible are biased towards a high $\ssigma$ in comparison to other galaxies of the same mass (which are also thought to host MBH). While this has little impact on the slope of the relations, this biases the normalisation of the $\mbh$-$\ssigma$ relation and the $\mbh$-$\mstar$ relation by up to a factor of 3. By contrast, biasing the sample to high masses artificially flattens the reported slopes of the scaling relations \citep{gultekin2009RELATIONSGALACTICBULGES}. This makes the lack of sampling in the IMBH regime particularly difficult \citep{greene2020IntermediateMassBlackHoles}. As the $\mbh$-$\ssigma$ and the $\mbh$-$\mstar$ relation are used to calibrate the other mass measurements (as well as cosmological simulations and semi-analytic models).

The scaling relations are not universal for all types of galaxies. As can be seen in Fig.~\ref{fig:scaling_relations}, BH in bulge-dominated galaxies (labelled as `early'-type galaxies in the legend) are fitted with different scaling relations than those in disc-dominated galaxies (`late'-type). However, given the continuous nature of galaxy morphology, one could also interpret this as a large scatter on the scaling relations depending on the evolutionary history of the galaxy. One difficulty in drawing causal conclusions form the different scaling relations is that there are often intrinsic differences between samples that might account for some of the differences. For example, the AGN plotted in the right panel of Fig.~\ref{fig:scaling_relations} have lower BH masses than predicted by scaling relations. 
Even from Fig.~\ref{fig:scaling_relations} it is easy to see that that as a consequence local AGN are preferentially hosted in lower-mass disc-dominated galaxies \citep{goulding2010CompleteCensusActive}, while bulge-dominated galaxies host quiescent MBH. This is confirmed when looking at AGN surveys which are dominated by disc-dominated galaxies \citep{koss2011HostGalaxyProperties,reines2015RELATIONSCENTRALBLACK}. On the scaling relations, bulge-dominated galaxies, quiescent MBH, unbarred galaxies and those with classical bulges, as well as stacked samples including all types of galaxies and MBH, have a steeper slope and tighter scatter than disc-dominated galaxies, AGN, bared galaxies and those with pseudo-bulges. This is likely because all of these categories preferentially probe SMBH in massive bulge-dominated galaxies. By contrast, AGN, disc-dominated galaxies and those with pseudo-bulges and bars have a shallower slope and larger scatter \citep{gultekin2009RELATIONSGALACTICBULGES,woo2013QUIESCENTACTIVEGALAXIES,saglia2016SINFONIBLACKHOLE,greene2020IntermediateMassBlackHoles}, mostly because all those categories preferentially select lower-mass galaxies. 

\begin{figure}
    \centering
    \includegraphics[width=\linewidth]{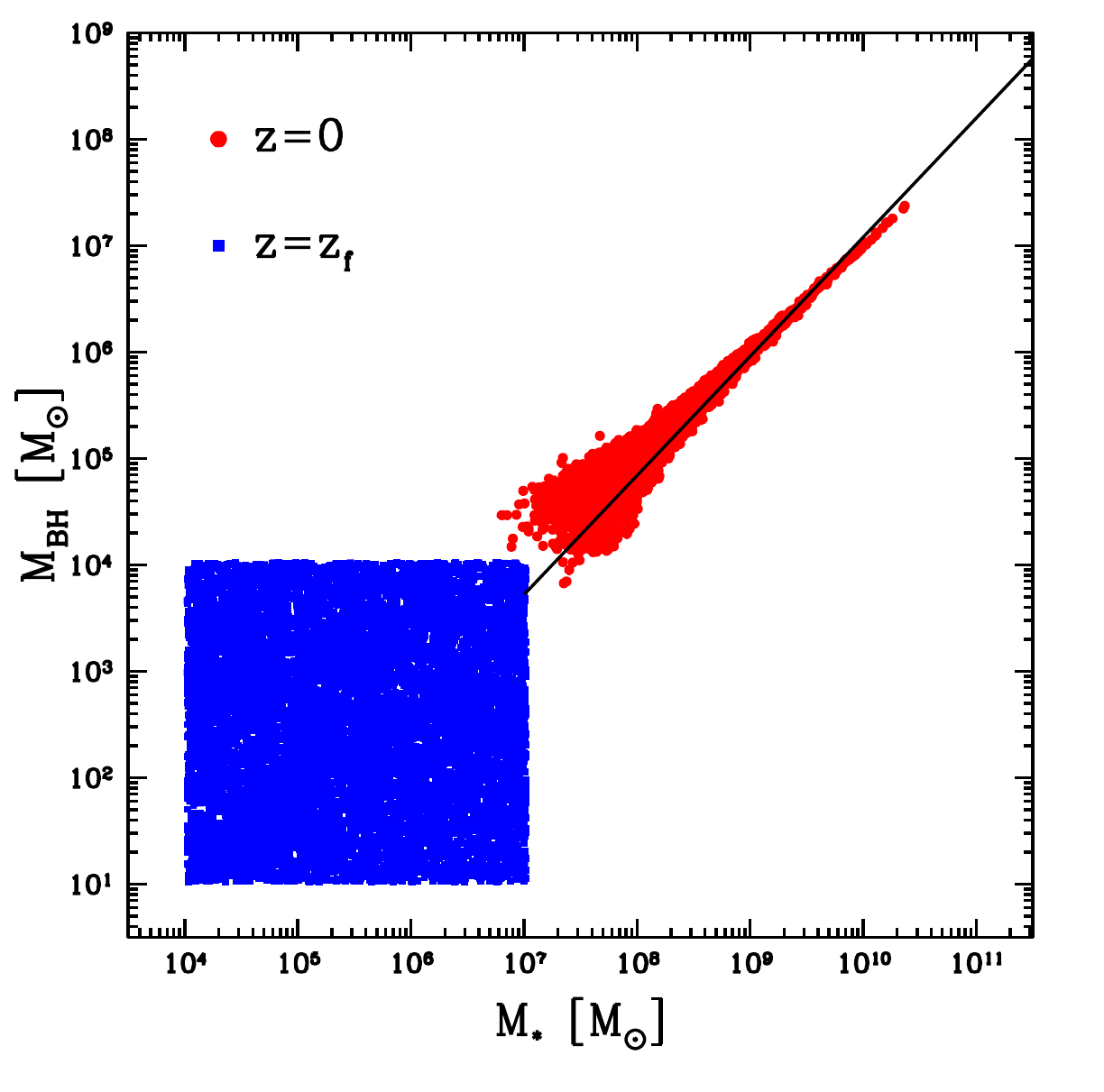}
    \caption{The local $\mbh-\mstar$ relation (red) emerges from repeated mergers along galaxy merger trees from an originally uncorrelated distribution at high redshift (blue). Reproduced with permission from \citet{jahnke2011NoncausalOriginBlackholegalaxy} }
    \label{fig:merger_averaging}
\end{figure}

These differences are interpreted as the existence of two sets of scaling relations, one for lower mass MBH that are currently actively growing, typically found in disc-dominated galaxies, and one for higher-mass MBH with low Eddington ratios that are now typically found in red bulge-dominated galaxies. The dichotomy between these populations suggests a fundamental difference in the processes driving the coevolution. In particular, the tightening scatter of the scaling relations for high masses, and high stellar velocity dispersion is thought to be due to the impact of galaxy mergers; it has been shown using semi-analytic models that repeated mergers of galaxies and MBH significantly reduce the scatter of the relation \citep{hirschmann2010EvolutionIntrinsicScatter,jahnke2011NoncausalOriginBlackholegalaxy} and that the number of past mergers increases as a function of mass (See Fig,~\ref{fig:merger_averaging}). 
However, galaxy mergers are thought to be rare events in the evolution history of each individual galaxy. Therefore, in contrast, the larger scatter of the scaling relations for lower mass, disk-dominated galaxies suggests a plethora of internal, so-called `secular', processes may be responsible for the coevolution in the epochs between galaxy mergers. Recent cosmological volume simulations predict that the coevolution of SMBH and galaxies is dominated by processes internal to galaxies (up to $85\%$ of SMBH growth since $z\sim3$; \citealt{martin2018, mcalpine2020}). Observational studies have attempted to isolate this growth pathway by selecting galaxies with merger-free evolutionary histories (i.e. those without the bulges grown by galaxy mergers), and have found that these galaxy-merger-free processes are still capable of reproducing galaxy-SMBH scaling relations \citep{ssl17}, and powering both MBH growth and AGN outflows \citep{Smethurst2021}.
However, galaxy-merger-free AGN are severely understudied. Previous observational studies either explicitly examine galaxies with merger-dominated histories \citep[e.g. ULIRGs;][]{tadhunter2018, perna2021} or use a mixed morphology sample of galaxies with muddled evolutionary histories (e.g. surveys such as CARS and 4MOST-AGN) where the effect of galaxy mergers will washout the effect of secular processes. %and by their nature also do not represent the most massive galaxies and MBH. 
Consequently, the physical processes responsible for the predicted {\it majority} of SMBH growth and AGN feedback are still poorly understood. We discuss this further in Section~\ref{sec:drivingBHgrowth}.

One way to test our understanding of coevolution is to use cosmological simulations. Cosmological simulations evolve numerical models for MBH growth and processes relevant for structure formation and galaxy evolution across cosmic time \citep[see][for a review]{vogelsberger2020CosmologicalSimulationsGalaxy}. We can compare the predicted synthetic populations of MBH and galaxies from such simulations to observations to understand whether our current model of MBH growth (see Sec.~\ref{sec:drivingBHgrowth}) could have produced the Universe we see. Current results look promising. Cosmological simulations predict MBH populations whose slope and normalisation (after tuning) match the observed $\mbh - \mstar$ and $\mbh-\ssigma$ relation well (see \citet{volonteri2016CosmicEvolutionMassive,mcalpine2018RapidGrowthPhase,weinberger2017SimulatingGalaxyFormation,pillepich2018SimulatingGalaxyFormation,dave2019SIMBACosmologicalSimulations} and \citet{habouzit2021SupermassiveBlackHoles} for a recent comparative analysis). They also predict the observed tightening of the MBH-galaxy scaling relations at high $\mbh$ but universally under-predict the scatter of the relations, especially for galaxies in the mass-range $10^{10} - 10^{11} \Msun$. It has been suggested that the lack of scatter could be due to the limited box sizes of simulations, which do not probe sufficiently rare environments to produce outlying objects. Another possible explanation for the lack of scatter is that current simulations merge MBH at un-physically large separations due to technical constraints. This effectively boosts the BH-BH merger rate and could lead to unphysical tightening of the relation through merger-averaging. If this is the case then current efforts to model the dynamical evolution of MBH binaries below the resolution scales of current simulations \citep{volonteri2020BlackHoleMergers,li2024TrackingOntheflyMassive}, which will lengthen merger timescales, should soon bring new insights. 

Any evolution in the scaling relations with redshift will give key insights into how MBH and galaxies coevolve and how the local scaling relations are established. If MBH are over-massive in the early Universe in comparison to galaxies this would suggest that MBH growth predates galaxy growth and it is the existence of a MBH that shapes late galaxy evolution. By contrast if galaxies are over-massive early on it is more likely that internal processes in the galaxy control MBH growth.

There is currently no evidence that the $\mbh-\mstar$ relation fundamentally changes with redshift out to at least $z=2$ \citep{cisternas2011SecularEvolutionNonevolving,schramm2013BlackHoleBulgeMass,sun2015EvolutionBlackHole}%, even though an evolution used to be reported \citep{merloni2009COSMICEVOLUTIONSCALING,decarli2010QuasarRelationCosmic,bennert2011RELATIONBLACKHOLE}.  
At higher redshift, the evolution is more uncertain. Local scaling relations report a typical specific MBH mass of around 0.01 \% \citep{reines2015RELATIONSCENTRALBLACK}. Recent observations from the James-Webb-Space telescope have reported that MBH in the early Universe ($4<z<11$) are over-massive in comparison to local scaling relations \citep{maiolino2024JADESDiversePopulation,harikane2023, stone2024UndermassiveHostGalaxies,natarajan2024}. It is currently difficult to ascertain whether this evolution is genuine or arises because the sample is biased towards extreme objects \citep{li2025TipIcebergOvermassive,jespersen2025SignificanceRareObjects}. It has been shown that such early MBH are less overmassive when compared to local quiescent MBH than to local AGN \citep{ellis2024ConsistencyJWSTBlack}, and also that overmassive MBH in the local Universe also preferentially live in compact host galaxies \citep{vandenbosch2012OvermassiveBlackHole} much like the high-redshift detections. Both suggest that there is little underlying evolution in the scaling relations. Cosmological simulations also predict little evolution of the $\mbh-\mstar$ relation but disagree on whether the remaining amount of evolution leads to an increase or a decrease in the normalisation of the scaling relations \citep{habouzit2022CoevolutionMassiveBlack}. Observations of faint quasars in the early Universe will be key to answering this question.

Another question raised by the scaling relations shown in Fig.~\ref{fig:scaling_relations}, is whether there is both a lower and upper limit to the mass of central MBH in galaxies. A broader set of observations confirms that maximum MBH masses appear to on the order of several $10^{10}$ \citep{Brockamp2016, Dullo2017, Ge2019, Dullo2021}, and in addition some of the highest redshift MBH have masses as high as the most massive MBH in nearby galaxies \citep{mortlock2011LuminousQuasarRedshift,wu2015UltraluminousQuasarTwelvebillionsolarmass,trakhtenbrot2011BLACKHOLEMASS, Whalen2024UHZ1}, which suggests that there might a theoretical maximum BH mass. This is hypothesised to be due to the lack of ability to grow by accretion for so-called `ultramassive' BH when the innermost stable circular orbit passes beyond the self-gravitational radius. At this point the MBH can no longer support an accretion disc (See Sec.~\ref{sec:accretion_states}). Depending on the BH spin, the upper mass limit is predicted to be between $10^{10.7}-10^{11.4}$ \citep{king2015HowBigCan},

The lower mass limit for central BH that potentially coevolve with their host galaxy is harder to constrain. 
The low mass of IMBH ($\lesssim10^6~\Msun$) means their sphere of gravitational influence for dynamical mass measurements is small \citep{baldassare2020PopulatingLowmassEnd}. Their colours are more difficult to differentiate from star formation than SMBH \citep{baldassare2018IdentifyingAGNsLowmass} and even when identified as AGN they are easily confused for X-ray binaries powered by highly-accreting stellar mass BH \citep{greene2020IntermediateMassBlackHoles}. To date, the lowest mass central BH was reported for the dwarf galaxy NGC 205 with a mass of just $M_{\rm BH} = 6.8 \times 10^3 \rm \ M_\odot$ \citep{nguyen2019ImprovedDynamicalConstraints}. Firm dynamical detections at this mass range are rare, with most measurements only reporting upper limits. From $ M_{\rm BH} = 10^5 \rm \ M_\odot$ dynamical detections become more common \citep{greene2020IntermediateMassBlackHoles} but the potential population of IMBH remains poorly sampled. 

To understand further how the observed scaling relations were established, we need to understand the growth of MBH across cosmic time which we can investigate directly by studying the changing distribution of AGN luminosities.

\section{Black hole growth across cosmic time}

MBH that are growing more rapidly emit stronger electromagnetic radiation. The bolometric luminosity of an AGN, $L_{\rm AGN}$ is thought to be directly proportional to the accretion rate of the central MBH, $\dot{M}_{\rm BH}$ as follows:
\begin{equation}
    L_{\rm AGN} = \epsilon_{\rm r} \dot{M}_{\rm BH} c^2
    \label{eq:LAGN}
\end{equation}
where $\epsilon_{\rm r} $ is the radiative efficiency, often assumed to be in the range $\epsilon_{\rm r}\sim0.07-0.15$, depending on the spin of the MBH \citep{Elvis2002}. MBH accretion rates can be parametrised using the Eddington accretion rate, 
\begin{equation}
\dot{M}_{\rm Edd} = \frac{4 \pi G M_{\rm BH} m_p} {\epsilon_{\rm r} c \sigma_T},
\label{eq:dotMedd}
\end{equation}
where $G$ is the gravitational constant, $M_{\rm BH}$ is the BH mass, $m_p$ is the proton mass, $c$ is the speed of light and $\sigma_T$ is the Thomson cross section. From this we define the Eddington ratio, $\lambda_{\rm Edd}$ is therefore $\dot{M}_{\rm BH} = \lambda_{\rm Edd} \dot{M}_{\rm Edd}$. High $\lambda_{\rm Edd}$ means the BH is accreting efficiently. Note that an inefficiently accreting MBH (with low $\lambda_{\rm edd}$ and high $M_{\rm BH}$) can have the same luminosity as an efficiently accreting low-mass MBH. 

It is not possible to directly measure the bolometric luminosity of an AGN; instead, corrections must be used to infer bolometric luminosities from narrow-band detections \citep{shen2020BolometricQuasarLuminosity}, most commonly the luminosity around \oiii 5007 in the optical, or in infrared wavebands. When inferring the bolometric luminosity one must also consider that attenuation from dust preferentially obscures some AGN, making it difficult to estimate the intrinsic luminosity of the MBH if infrared observations are not available. In addition it is worth noting that bolometric corrections are not constant and can depend on BH mass \citep{heckman2014CoevolutionGalaxiesSupermassive} and Eddington ratio \citep{vasudevan2007PiecingTogetherXray, vasudevan2009OpticalXrayEmissionLowabsorption}. Despite these difficulties, studying AGN luminosity gives us the most direct access to track the coevolution of galaxies and their SMBH across cosmic time.

Observations of AGN in the local Universe show that there are several categories of AGN that have significantly different spectral properties. After much work it was recognised that at least some of the differences are due to viewing angle \citep{beckmann2012ActiveGalacticNuclei}. We refer the reader to the chapter on AGN for further details. However, some of the observed differences in observed AGN are driven by an underlying shift in the structure of the accretion on the MBH. In this section we first discuss the physics of two accretion modes, before discussing observed MBH growth in the local universe and across cosmic time in the context of the two growth modes. 

\subsection{Black hole accretion modes}
\label{sec:accretion_states}
At a threshold Eddington ratio of $\fedd = 0.01$ the structure of the accretion disc around the BH significantly changes. This impacts the radiative efficiency $\epsilon_{\rm r}$ as well as the form in which the feedback energy of the MBH is being ejected. This shift in accretion is independent of BH mass, and is thought to be universal from stellar mass BH to SMBH. 

At high Eddington ratios ($\fedd \gtrsim 0.01$) the accretion disc around the BH is geometrically thin and optically thick \citep{shakura1973BlackHolesBinary}. The energy output of such AGN is dominated by photons, as the radiative efficiency of such thin discs is high. A significant fraction of the accreted energy is released in the form of a blackbody spectrum from the inner part of the disc, which is also the hottest. The radius of this inner edge, the innermost stable orbit, determines the effective temperature of the emitted spectrum and therefore the radiative efficiency, $\epsilon_{r}$, depends on the BH spin \citep{bardeen1972RotatingBlackHoles}. The highest radiative efficiencies are found for MBH that have the maximum permissible spin of $a = 0.998$ where $a$ is the dimensionless spin parameter of the BH \citep{thorne1974DiskAccretionBlackHole}. Some of this radiation is reprocessed in the broad-line region and the narrow-line region, adding both broad and narrow emission lines to the AGN spectrum. Observationally not all highly accreting radiatively efficient AGN look the same, as a dusty obscuring structure can absorb or block specific parts of the spectrum for some AGN, depending on viewing angle \citep[see][for a review]{beckmann2012ActiveGalacticNuclei}. As the accretion disc is optically thick, it absorbs some of its own radiation which drives hot, wide-angle winds that impact the interstellar medium (ISM) of the host galaxy. We will refer to radiatively efficient AGN as being in ``quasar mode" (although note that this is also referred to as ``thermal mode" in the literature). 

At lower Eddington ratios ($\fedd \lesssim 0.01 $), the inner accretion disc transitions to a geometrically thick, optically thin structure, which fundamentally changes the behaviour of the AGN. While it is likely that the outer accretion disc remains thin, the overall radiative efficiency of the disc significantly drops, which means much less energy is being emitted in the form of photons. The optically thin accretion disc fails to capture this radiation and disc winds cease. Instead, energy output of the AGN is dominated by highly collimated, kinetic jets launched from the inner edge of the accretion disc \citep{blandford1977ElectromagneticExtractionEnergy} which can be detected through their radio emission. Such jets are thought to be powered by the spin energy of the central BH and their efficiency is also higher for highly spinning BH. In radiatively inefficient AGN, broad lines are absent but weak narrow lines can sometimes be found. We will refer to such radiatively inefficient AGN as being in "jet mode" (although note that this is also referred to as ``radio mode" in the literature). Note that the picture of MBH accretion modes presented here has been simplified for the purpose of this discussion. This includes the simplification of the impact of the shift in accretion disc structure on the observable properties. For example, some "quasar mode" AGN are also observed to have jets. We refer readers to \citet{heckman2014CoevolutionGalaxiesSupermassive} and the chapter on AGN for further details. 

While MBH accrete in both modes, significant mass growth only occurs during quasar mode accretion (or through BH-BH mergers). The characteristic timescale for MBH growth is the effective Salpeter time, 
\begin{equation}
\tau_{\rm Sal} = \frac{\mbh}{\fedd \dot M_{\rm Edd}} \sim 50 \fedd^{-1} \rm \ Myr 
\label{eq:salpeter}
\end{equation} 
where $\dot{M}_{\rm Edd}$ is the Eddington luminosity from Eq.~\ref{eq:dotMedd}. The mass of a MBH accreting at constant Eddington ratio $\fedd$ is exponential and will grow over time $t$ as $\mbh(t) = M_{\rm BH,0} \exp{t / \tau_{\rm Sal}(f_{\rm Edd})}$ where $M_{\rm BH,0}$ is the initial MBH mass. This means that over a Salpeter time, a MBH accreting at a constant Eddington ratio, $\fedd$, will increase its mass by a factor of 2.7. As can be seen from Eq.~\ref{eq:salpeter}, for MBH in jet mode the effective Salpeter time quickly exceeds the age of the Universe. By contrast, as $\fedd \rightarrow 1$, or especially when $\fedd > 1$ during so-called ``super-Eddington" accretion, MBH grow significantly in mass in short timescales. Note that accretion disc simulations have shown that BH can easily achieve super-Eddington accretion rates if fed at sufficiently high rates from the galaxy \citep{inayoshi2020AssemblyFirstMassive}, providing in principle a formation pathway for SMBH observed at high redshift \citep{schneider2023AreWeSurprised}. There are observations of potentially Super-Eddington MBH in the early Universe \citep{trakhtenbrot2011BLACKHOLEMASS,lupi2024SizeMattersAre}. Unfortunately, cosmological simulations suggest that the conditions for sustained super-Eddington accretion are rare even in the early Universe \citep{regan2019SuperEddingtonAccretionFeedback,massonneau2023SuperEddingtonGrowth} and that allowing for Super-Eddington accretion can actually stunt MBH growth in the long run due to strong feedback effects. 

\subsection{Massive black hole growth in the local Universe}
\label{sec:MBHgrowthlocal}

\begin{figure}
    \centering
    \includegraphics[width=\textwidth]{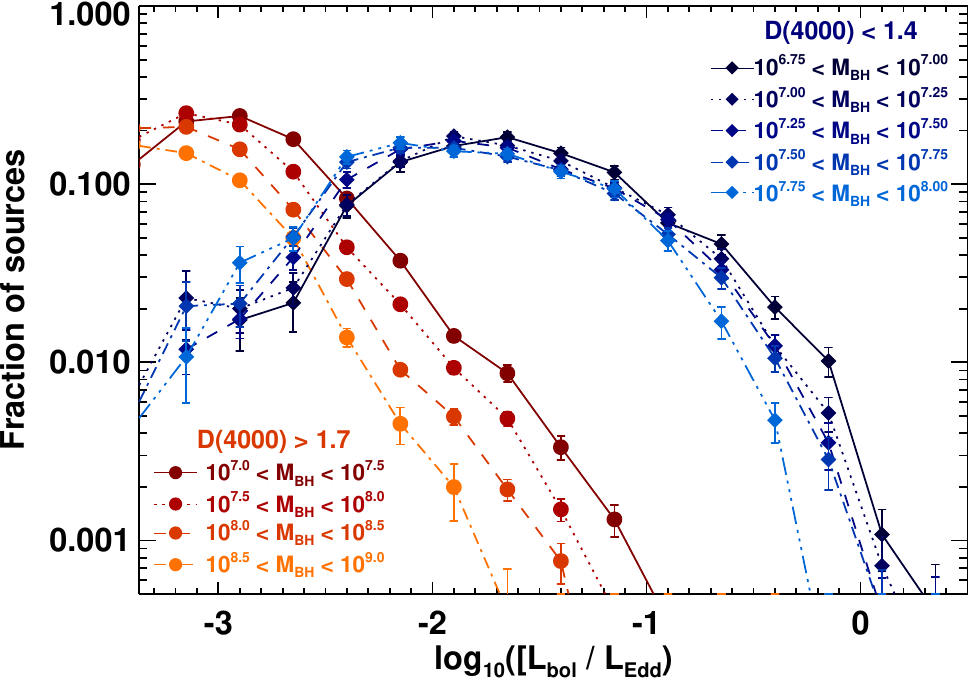} 
    \caption{Distribution of Eddington ratios $\fedd$ for AGN in the local Universe split into star-forming galaxies (blue) and quenched galaxies (red) from \citet{heckman2014CoevolutionGalaxiesSupermassive}. More massive BH (lighter colours) have lower Eddington ratios for both populations. Modified with permission from the Annual Review of Astronomy and Astrophysics, Volume 58 © 2020 by Annual Reviews, http://www.annualreviews.org}
    \label{fig:Eddington}
\end{figure}

In the local Universe, both accretion modes are present but they are not uniformly distributed with BH mass. The highest Eddington ratios are found for MBH in the mass range $10^6-10^8 \rm  \ \Msun$ which accrete in the radiatively efficient quasar mode \citep{heckman2004PresentDayGrowthBlack}. Peak Eddington ratios for radiatively efficient AGN in the local Universe rarely exceed $\fedd = 0.1$, with 99.8 \% of AGN having Eddington ratios smaller than that (see Fig.~\ref{fig:Eddington}) \citep{heckman2004PresentDayGrowthBlack}. When studying AGN in the local Universe, it is MBH in this mass range which dominate the sample \citep[e.g.][]{reines2015RELATIONSCENTRALBLACK}. Local Eddington ratios fall with increasing MBH mass (Lighter colours in Fig.~\ref{fig:Eddington} indicate higher MBH mass). The most massive MBH on average have the lowest Eddington ratios \citep{heckman2004PresentDayGrowthBlack} and as a result are almost uniformly thought to be in jet mode $\fedd < 0.01$ \citep{heckman2004PresentDayGrowthBlack}. Current surveys show no sign of the distribution of Eddington ratios tailing off at the low brightness end \citep{kauffmann2009FeastFamineRegulation}.

The distribution of $\fedd$ is difficult to constrain because AGN become harder to detect as their Eddington ratios drops. Due to the degeneracy between $\mbh$ and $\fedd$, flux-limited samples can be incomplete across all mass bins. This also makes it difficult to study the distribution of Eddington ratios, especially at the low $\fedd$ end, even for clearly defined mass bins \citep{kelly2008ObservationalConstraintsDependence}. For a given AGN survey, one can estimate the BH masses for all AGN above the luminosity threshold using single-epoch virial mass estimators. Once masses have been determined, Eddington ratios can be estimated from the bolometric luminosity. A more accurate way is to simultaneously fit for both $\mbh$ and $\fedd$ using templates of AGN spectra, as done e.g. in \citet{trakhtenbrot2011BLACKHOLEMASS}. It is possible to estimate MBH mass functions, and their Eddington ratios, for the whole population of MBH using a population synthesis approach. To do so, one assumes a MBH mass function, creates synthetic fluxes for that mass function, applies incompleteness and selection functions and then uses a Bayesian approach on observed luminosity functions to constrain the models \citep{kelly2009DeterminingQuasarBlack,merloni2008SynthesisModelAGN}.

As shown in Fig.~\ref{fig:Eddington}, efficient MBH growth in the local Universe is associated with star formation: Eddington ratios are higher in star-forming galaxies than in those with little ongoing star formation. Star forming galaxies are more likely to host AGN \citep{azadi2015PRIMUSRELATIONSHIPSTAR} at a given stellar mass. The correlation is not linear as the MBH growth efficiency saturates for high star formation rates \citep{kauffmann2009FeastFamineRegulation,azadi2015PRIMUSRELATIONSHIPSTAR}. \citet{heckman2014CoevolutionGalaxiesSupermassive} showed that Eddington ratios are different distributed for starforming and quenched galaxies: a log-normal distribution for MBH in star-forming galaxies (blue in Fig.~\ref{fig:Eddington}), which includes all of the most efficiently growing AGN in the local Universe, and a power-law slope for MBH in quiescent galaxies (red in Fig.~\ref{fig:Eddington}). The two populations roughly map to the two accretion modes. Star forming galaxies preferentially host radiatively efficient AGN in quasar mode, while quiescent galaxies hosting less efficient AGN in radio mode. 

The host galaxies of these radiatively efficient present-day AGN are thought to be evolving secularly. One way to differentiate secularly evolving spiral galaxies from those that have in the past undergone a major merger is through the structure of the central bulge: a pseudo-bulge with a flattened Sersic index indicates the dominance of secular evolution, while a classical bulge with a steeper Sersic index is a clear indication of past merger activity \citep{fisher2008STRUCTURECLASSICALBULGES}. There is growing evidence in favour of the dominance of secular MBH growth and coevolution, including recent surveys reporting pseudo bulges in the majority of local AGN hosts \citep{jiang2011BLACKHOLEMASS, malkan1998HubbleSpaceTelescope}, the energetics of AGN outflows in galaxy-merger-free hosts being statistically indistinguishable from the local AGN population \citep{Smethurst2021}, the high Eddington ratios of AGN hosted by merger-free galaxies \citep{ssl17}. Further evidence comes from the fact that most local AGN are located in galaxies without signs of morphological disturbance that would suggest a recent interaction or merger \citep{cisternas2011BulkBlackHole,schawinski2012HeavilyObscuredQuasar}. While classical bulges dominate the number density of samples \citep{gadotti2009GrowthSupermassiveBlack}, due to their lower Eddington ratios they only contribute approximately 10 \% of BH mass growth in the local Universe \citep{wild2007BurstyStellarPopulations}. We will discuss the potential impact of galaxy mergers on BH growth further in Section~\ref{sec:merger_growth}.

The observational evidence there suggests that the majority of MBH growth in the local Universe occurs in moderately massive, star forming, secularly evolving spiral galaxies onto relatively lower mass SMBH \citep{wild2007BurstyStellarPopulations,aird2012PRIMUSDependenceAGN} (although note that \citet{ssl17} measured MBH masses up to $10^9~\rm{dex}$ in a sample of secularly powered AGN). Typically though, the most massive local SMBH have completed their mass growth through gas accretion and are now found in a low accretion state in equally quiescent galaxies. 

\subsection{The evolution of massive black hole growth across cosmic time}

\begin{figure}
    \centering
    \includegraphics[width=\linewidth]{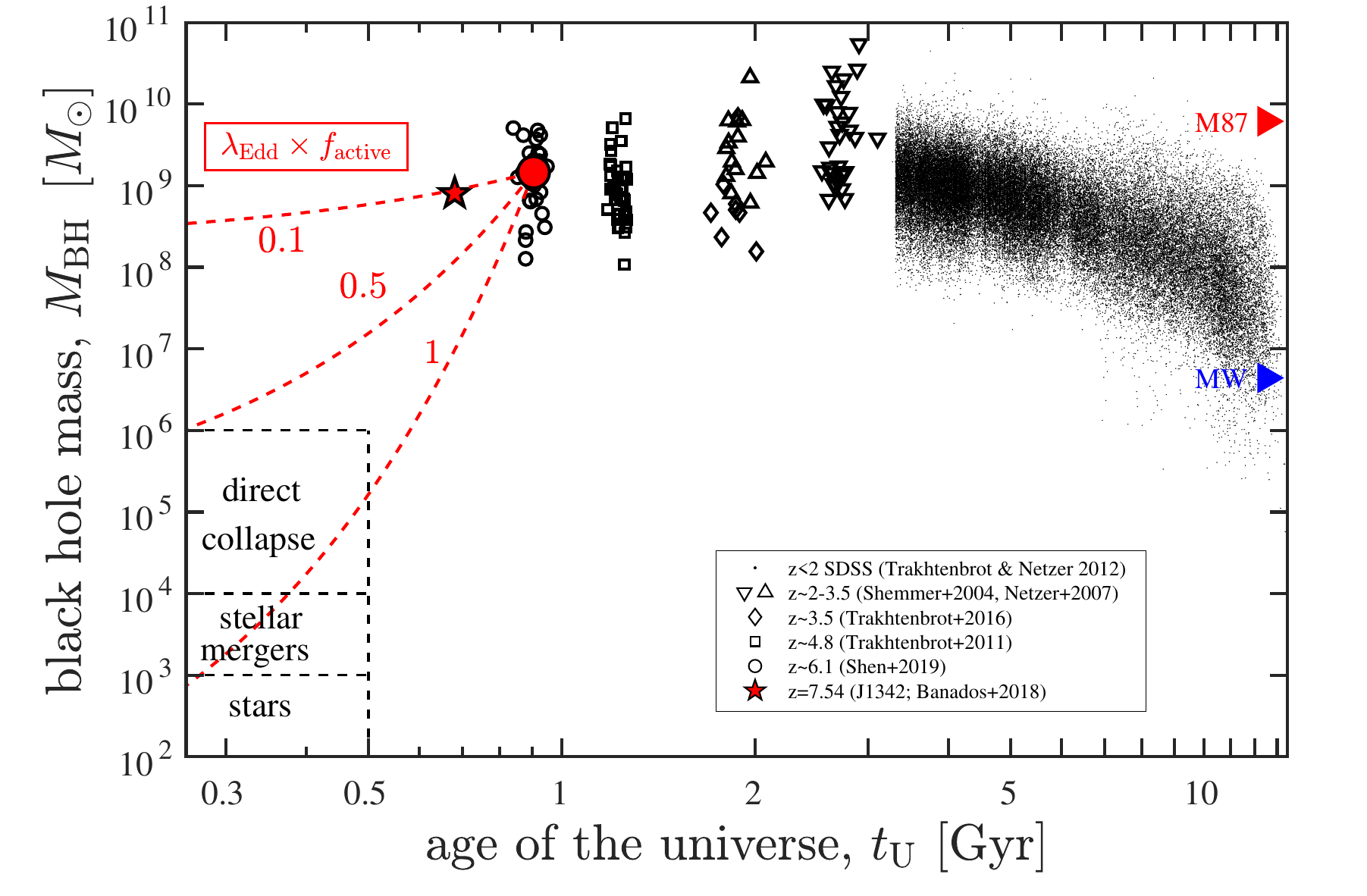}
    \caption{Time evolution of the MBH masses measured for AGN from cosmological surveys (black and red markers). Average MBH masses of active BH increased until cosmic noon ($z=2$). Cosmic downsizing describes the phenomenon shown here that the mass of MBH powering AGN has been decreasing since $z=2$ ($t \approx 3 \rm \ Gyr$). As indicated by the red dashed lines, seed black holes (rectangular boxes) must have grown very effectively since their formation to reach the high observed MBH masses in the early Universe. Courtesy of B. Trakhtenbrot based on \citet{trakhtenbrot2021WhatObservationsTell}}
    \label{fig:AGN_downsizing}
\end{figure}

% \begin{figure}
% 	\centering
% 	\includegraphics[width=\linewidth]{Figures/trakhtenbrot2012BlackHoleGrowth_fig16.pdf}
% 	\caption{Redshift evolution of the Eddington ratio $\fedd$ for MBH in the mass range $10^{7.5-7.8}\Msun$ (red), $10^{8.5-8.8}\Msun$ (blue) and $10^{9.5-9.8}\Msun$ (green) as a function of time. By $z=2$, accretion onto the most massive BH had already become less efficient. Over time, the mass of the most active MBH decreased. Reproduced with permission from \citet{trakhtenbrot2012BlackHoleGrowth}.\textcolor{red}{PERMISSION PENDING}}
% 	\label{fig:BHgrowth_historic}
% \end{figure}

Given that the first AGN have been observed as early as $z=11$ \citep{maiolino2024SmallVigorousBlack,bogdan2024EvidenceHeavyseedOrigin, Whalen2024UHZ1} when the Universe was less than a billion years old, BH growth must span the majority of cosmic history. As can be seen in Fig. \ref{fig:AGN_downsizing}, the SMBH powering some of the first observed high redshift AGN already had masses comparable to present-day MBH \citep{mortlock2011LuminousQuasarRedshift,wu2015UltraluminousQuasarTwelvebillionsolarmass,trakhtenbrot2011BLACKHOLEMASS}, so the onset of BH mass growth dates back even further than our earliest observations. Observations become technically more challenging and the minimum required luminosity for an AGN to be observed is much higher in the early Universe. Both effects together mean that the higher the redshift the less representative our observed sample of AGN are likely to be of the total population of MBH. This is confirmed by the fact that the average MBH mass of AGN in Fig.~\ref{fig:AGN_downsizing} increases until $z=2$.

One approach to studying the evolution of MBH growth over time is to look at the total BH mass accretion density as a function of time. The BH mass accretion density is the total mass accreted onto all MBH per unit volume at a given redshift. It can be found by integrating the bolometric AGN luminosity function, assuming some value for the radiative efficiency, $\epsilon_{\rm r}$. Observational studies have found that this accretion rate density of MBH increases with cosmic time until a peak at cosmic noon ($z\sim 2$) \citep{shankar2009SELFCONSISTENTMODELSAGN,aird2010EvolutionHardXray} and has been declining ever since. This closely tracks the evolution of the star formation density \citep{hopkins2004EvolutionStarformingGalaxies,hopkins2006NormalizationCosmicStar,fardal2007EvolutionaryHistoryStars}, although there is tentative evidence that the BH accretion density peaks before the star formation density. Integrating the BH mass formation density gives a surprisingly good match to the estimated BH mass density in the local Universe \citep{mclure2004CosmologicalEvolutionQuasar}, which shows that the bulk of BH mass was assembled through radiatively efficient accretion. Note that this predominantly constrains the growth of massive SMBH as they dominate the BH mass budget, despite the fact that low-mass SMBH and IMBH dominate the number density. 

%The distribution of Eddington ratios, and the masses of the most active MBH, has also evolved strongly with time, as can be seen in Fig.~\ref{fig:BHgrowth_historic}. 
At least since $z=5$, MBH growth has become less efficient as average Eddington ratios have been decreasing \citep{netzer2007BlackHoleMass,trakhtenbrot2011BLACKHOLEMASS,trakhtenbrot2012BlackHoleGrowth}. At all redshifts, the distribution of Eddington ratios remains broad across all MBH masses, and the density of MBH continues to increase towards Eddington ratios below survey completeness limits \citep{merloni2008SynthesisModelAGN, kelly2012MassFunctionsSupermassive}. This broad evolution of Eddington ratio at all MBH masses shows that MBH growth has a duty-cycle, where periods of efficient growth are interspersed with periods of quiescence. Estimates suggest that most MBH roughly double their mass from $z=5$ to $z=2$. Over this epoch of $ \sim 1.5 \rm \ Gyr$, they are estimated to be active 10-20 \% of the time to match observed luminosity functions \citep{yu2002ObservationalConstraintsGrowth,mclure2004CosmologicalEvolutionQuasar,haiman2004ReasoningFossilsLearning,kelly2010CONSTRAINTSBLACKHOLE,trakhtenbrot2011BLACKHOLEMASS}. 

In a phenomenon known as "cosmic downsizing of AGN``, the mass of the most active MBH has been decreasing since $z=2$  \citep{marconi2004LocalSupermassiveBlack, merloni2004AntihierarchicalGrowthSupermassive, trakhtenbrot2012BlackHoleGrowth} with the Eddington ratio of the most massive BH dropping faster than of less massive BH (see Fig.~\ref{fig:AGN_downsizing}). By $z=2$, the most massive BH had already stopped growing as efficiently than lower-mass BH suggesting that the peak of MBH growth occurred at even higher redshift.

To grow in mass through radiatively efficient accretion, MBH need to be fed by ample cosmic gas. A quick estimate shows that the hot interstellar medium is insufficiently dense to sustain the observed accretion rates for the vast majority of AGN in the local Universe \citep{bondi1952SphericallySymmetricalAccretion,allen2006RelationAccretionRate}. It has been suggested that some low-excitation radio sources, found in gas-poor massive elliptical galaxies, are fed directly from the hot intra-galactic medium \citep{hardcastle2007HotColdGas, allen2006RelationAccretionRate}. It was later shown that while the energetics balance, the correlation between jet feedback power and the estimated accretion rates in those sources is too weak to support the argument \citep{russell2013RadiativeEfficiencyVariability}. The fact that MBH are not fed from the hot interstellar medium is also supported by observations that show a lack of correlation between the amout of observed atomic gas in the host galaxy and the AGN activity \citep{fabello2011AreciboLegacyFast}. Instead, 
AGN in the local Universe \citep{heckman2014CoevolutionGalaxiesSupermassive} and out to $z=5$ \citep{suh2019MultiwavelengthPropertiesType} are observed to be located in galaxies with an abundant cold dense (molecular) gas supply, which is centrally concentrated \citep{hicks2013FUELINGACTIVEGALACTIC}.
 
Galaxy-scale and cosmological simulations model MBH accretion by assuming that a fraction of the local gas supply near the BH at the resolution scale of the simulation is accreted by the MBH. All schemes currently used, from a flux-based scheme for very high resolution simulations, via the Bondi-Hoyle-Lyttleton accretion rate \citep{bondi1952SphericallySymmetricalAccretion,edgar2004ReviewBondiHoyleLyttletonAccretion} to the torque-based accretion scheme proposed by \citet{hopkins2011AnalyticModelAngular} are designed to compute higher accretion rates for MBH environments rich in dense, cold gas. Most simulations also assume that a fraction of this gas is returned as feedback energy to the local MBH environment, following Eq.~\ref{eq:LAGN}, and that MBH can grow through BH-BH mergers. While the overall scheme of MBH growth and feedback is similar across most recent simulations, the details and model parameters vary. We can use a comparison of simulation predictions to identify robust features of our current model, as well as identify avenues for improvement due to their differences. It should be noted that free parameters in the accretion and feedback model of all simulations are calibrated for simulation populations to match at least one of the local MBH-galaxy scaling relations, as well as usually the galaxy stellar mass function and star formation rate density \citep[see][for a comparative discussion]{habouzit2022CoevolutionMassiveBlack}. When using simulations to assess the robustness of our model of MBH growth, it is therefore important to focus on measures not used for calibration, or to focus on remaining discrepancies with observations.

\begin{figure}
    \centering
    \includegraphics[width=\linewidth]{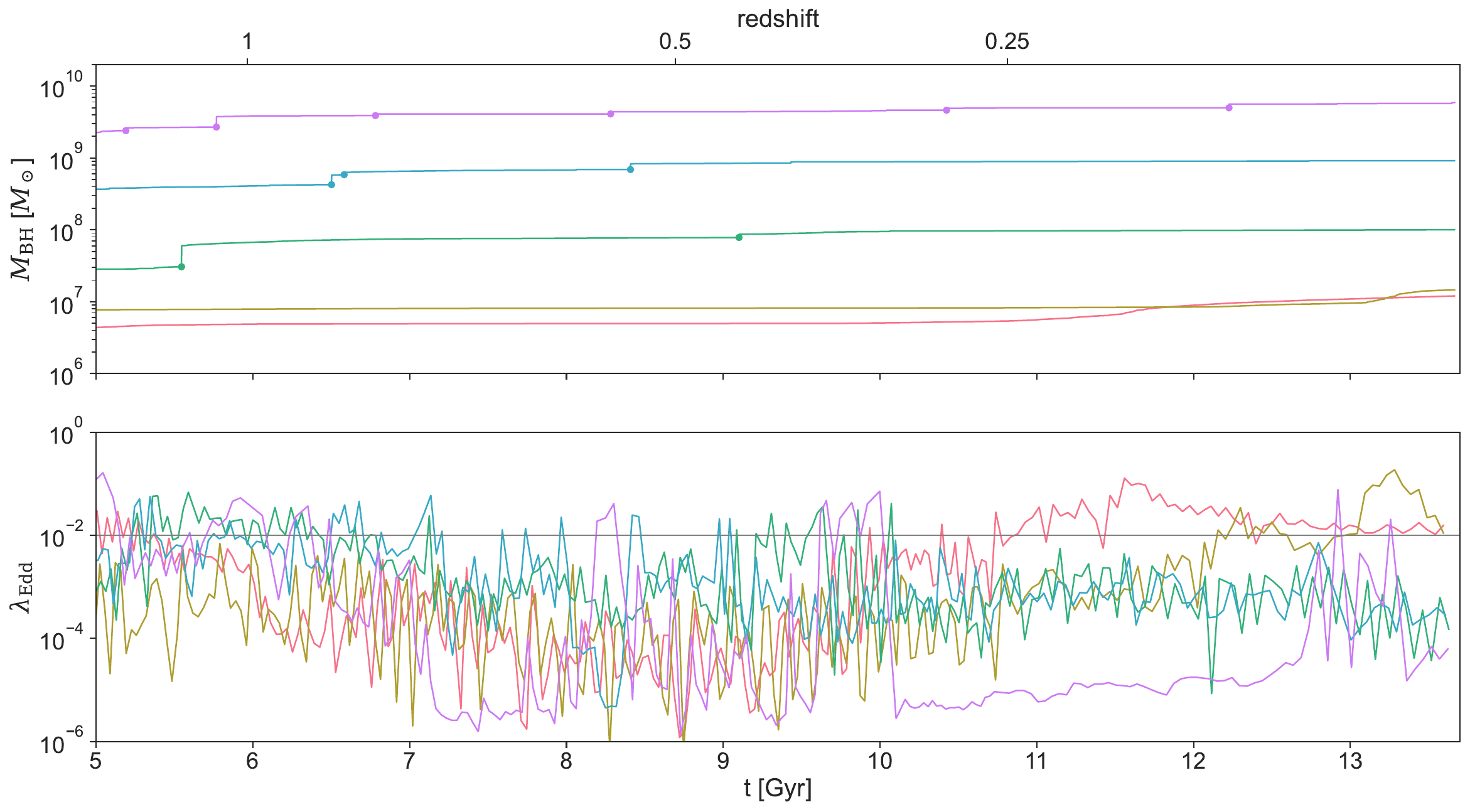}
    \caption{MBH evolution histories across cosmic time for five example MBH showing the MBH mass evolution (top panel) and the evolution of the Eddington ratio $f_{\rm Edd}$ (bottom panel) since $z=2$ for five MBH from the Horizon-AGN simulation  \citep{dubois2014DancingDarkGalactic,volonteri2016CosmicEvolutionMassive}. The Eddington ratio has been averaged over 100 simulation timesteps for legibility. Markers in the top panel denote BH-BH mergers. This plot shows that MBH accretion shows long-term evolution over Gyr timescales where individual AGN undergo periods of activity and quiescence. There is also variability in accretion efficiency on much shorter timescales. In this plot, we can see cosmic downsizing of AGN in progress. At $z<0.25$, the most massive SMBH (purple) also has the lowest $\fedd$, while the lowest-mass MBH are or are beginning to accrete the most efficiently. }
    \label{fig:MBH_growth_histories}
\end{figure}

Cosmological simulations using the cold gas based, two-mode accretion and associated feedback presented in Sec.~\ref{sec:accretion_states} reveal that MBH growth for any given MBH varies strongly over time. Simulations have shown that early MBH growth is suppressed and erratic in the early Universe because the cold gas supply in gas-rich, low-mass galaxies is controlled by supernova feedback from stars which heat and expel gas \citep{dubois2015BlackHoleEvolution,angles-alcazar2015TORQUELIMITEDGROWTHMASSIVE,prieto2017HowAGNSN,habouzit2017BlossomsBlackHole,beckmann2023PopulationStatisticsIntermediatemass,koudmani2021LittleFABLEExploring}. This significantly hampers MBH growth and leads to low Eddington ratios early on. It also produces a break in the $\mbh-\mstar$ relation at low $\mbh$, which is tentatively beginning to be ruled out by observations of IMBH \citep[see][for a recent review]{greene2020IntermediateMassBlackHoles}. This could be evidence that our current models might over-estimate the strength of stellar feedback in dwarf galaxies \citep{koudmani2022TwoCanPlay}.  

Around redshift $z=2-3$, simulations show that the galaxy is sufficiently massive for the gas supply to settle down and the simulated MBH enter a phase of efficient growth \citep{dubois2015BlackHoleEvolution,angles-alcazar2015TORQUELIMITEDGROWTHMASSIVE,prieto2017HowAGNSN,habouzit2017BlossomsBlackHole,koudmani2021LittleFABLEExploring}, in agreement with observations that show the peak of MBH growth during this period. At late times, simulations naturally predict the AGN downsizing seen in observations. Fig.~\ref{fig:MBH_growth_histories}, which shows the late accretion history for five representative MBH in the mass range $10^7-10^{10}$ from the Horizon-AGN simulation \citep{dubois2014DancingDarkGalactic,volonteri2016CosmicEvolutionMassive}, shows that the more massive BH, the earlier the onset in the fall of $\fedd$ and the lower the average $\fedd$ over the last Gyr. Such cosmological simulations have shown that accretion onto MBH is highly variable so the active periods of MBH are likely broken into a series of repeated short active phases, the so-called ``duty cycle", which can be as short as several Myr \citep{dubois2015BlackHoleEvolution}. Some of this variability might be smoothed out if accretion discs were modelled more explicitly in cosmological simulations, as they could act as a long-term, slow-release gas reservoir for the black hole. However, observations from the local Universe also support intermittent periods of AGN activity on timescales similar to those reported in current simulations \citep{schawinski2015ActiveGalacticNuclei}, and the lack of observed correlations between overall galactic gas supply and AGN activity also suggest that AGN activity could be highly intermittent.  There is also evidence for longer evolution cycles of MBH activity on Gyr timescales which can vary strongly even for MBH in the same mass bin (compare the two lowest mass MBH in Fig.~\ref{fig:MBH_growth_histories} who enter their latest active phase over 2 Gyr apart). Even otherwise inactive MBH can become active due to changes in their environment (see e.g. the most massive MBH around redshift $z=0.25$ in Fig.~\ref{fig:MBH_growth_histories}) We discuss the possible origins of such intermittent growth further in Section~\ref{sec:drivingBHgrowth}.

Semi-analytic models and cosmological simulations have shown that while the two-mode BH growth model with associated feedback generally reproduces observed MBH populations well, there are a few key challenges that will provide avenues for further investigation. One issue is the redshift evolution of MBH masses. Semi-analytic models in particular struggle to reproduce the observed early SMBH without over-predicting masses of SMBH in the local Universe \citep{natarajan2012MassFunctionBlack}. Galaxies in cosmological simulations consume their cold gas earlier than in semi-analytic models  \citep{hirschmann2012GalaxyFormationSemianalytic} and struggle less to regulate MBH masses at late times, provided MBH growth is allowed to self-regulate via AGN feedback, or computed based on large-scale torques in the galaxy \citep{angles-alcazar2013BLACKHOLEGALAXYCORRELATIONS, catmabacak2022BlackHolegalaxyScaling}. This shows that the clumpy distribution of cold gas in galaxies, and its response to AGN and stellar feedback injections, are one key driver in the coevolution of MBH and galaxies. We discuss this further in Sec.~\ref{sec:impact_on_galaxies}.

One challenge of this model of self-regulating growth is that the overall growth of MBH across cosmic time is limited. This puts a lower limit on the seed masses of MBH. ``BH seeds" refers to the populations of BH that grew into the MBH that we see today. Proposed BH seed formation mechanisms broadly fall into two categories: Light seeds are thought to be primarily stellar remnants and will have masses at formation of $< 10^3 \rm \Msun$. Heavy seeds already fall into the IMBH category at formation, with initial masses in the range $10^3 - 10^6 \rm \ M_\odot$ form either from run-away mergers in stellar clusters or from direct collapse \citep{volonteri2010FormationSupermassiveBlack,inayoshi2020AssemblyFirstMassive,volonteri2021OriginsMassiveBlack,regan2024MassiveBlackHole}. Light seeds are far more abundant than heavy seeds but face the challenge of having to grow by many more orders of magnitude to reach observed masses. This means their duty cycles need to be much longer, especially in the early Universe. As shown in Fig. \ref{fig:AGN_downsizing}, light seeds have to grow continuously at the Eddington limit to reach the observed masses. Heavy seeds can be active for shorter periods of time but they still need to grow efficiently for many Salpeter times to become present-day SMBH. Proposed mechanisms for heavy seed formation rely on very specific conditions which means heavy BH seeds are expected to be much rarer. It should also be noted that while we have abundant evidence for stellar mass BH \citep{abbott2023GWTC3CompactBinary}, the formation of heavy seed BH is much less well observationally supported. One problem is that the mechanisms for heavy seed formation all require very specific conditions which can only have been met in the early Universe. It has been claimed that UHZ1 might be a recently formed heavy seed BH but the claim remains contested \citep{bogdan2024EvidenceHeavyseedOrigin}. 

Mass growth through both gas accretion and BH-BH mergers is only efficient in the centre of galaxies, where there is an abundant dense, cold gas supply, and where BH can form close binaries in order to merge \citep{volonteri2020BlackHoleMergers}. Recent simulation case studies have highlighted how difficult it is for BH to find their way to the centres of low-mass galaxies, as orbits of seed BH in rapidly-evolving high redshift galaxies are complex\citep{bellovary2019MultimessengerSignaturesMassive,ma2021SeedsDontSink}, timescales for BH to sink to the centres of galaxies are long\citep{pfister2019ErraticDynamicalLife} and BH are easily dislodged again from early galactic centres \citep{bellovary2021OriginsOffcentreMassive}, confirming earlier analytic results \citep{bellovary2019MultimessengerSignaturesMassive,volonteri2005RapidGrowthHighRedshift}. Understanding the dynamics of MBH in galaxies, and the source of BH seeds of present-day MBH, remains two of the ongoing challenges in the field. The seed mass significantly influences the slope of the low-mass end of the $\mbh-\mstar$ relation \citep{habouzit2017BlossomsBlackHole} as well as the occupation fraction of dwarf galaxies at low redshift \citep{volonteri2008EvolutionMassiveBlack}. Simulations have shown that both seed types need to be in the centres of galaxies for long periods of time to grow effectively. To produce the first observed MBH, light seeds will have to grow at Super-Eddington rates at least some of the time \citep{madau2014SUPERCRITICALGROWTHMASSIVE}. 

Remaining discrepancies between simulated and observed MBH populations provide an interesting opportunity to further develop our model of MBH growth. One area that remains a challenge for current cosmological simulations are the observed masses of the first SMBH in the early Universe, which are frequently absent in simulated MBH populations. This is partially due to the rarity of these early SMBH. Simulations that cover a significantly larger volume than average have reported the formation of very massive early SMBH in agreement with observations \citep{tenneti2019TinyHostGalaxy}. More broadly the expanding observed population of SMBH at higher and higher redshift provides a challenge to the supernova-regulated early MBH growth model, in which early MBH growth lags that of the host galaxy stellar mass, that is favoured by current simulations. It is difficult to understand how it allows for MBH to be overmassive in early, low-mass galaxies. Similar questions concern IMBH in dwarf galaxies. Current simulations that have the required resolution do predict IMBH in dwarf galaxies but  under-predict the active fraction in local galaxies \citep{koudmani2021LittleFABLEExploring,beckmann2023PopulationStatisticsIntermediatemass}. There is also mounting observational evidence that  break in the $\mbh-\mstar$ relation predicted by current cosmological simulations based on heavy seeds and super-nova controlled early MBH growth is not reflected in observations \citep{greene2020IntermediateMassBlackHoles}. Both frontiers point to the fact that our current model of IMBH growth and their coevolution with their dwarf host galaxies might need reconsidering.

\section{Fuelling massive black hole growth} 
\label{sec:drivingBHgrowth}

To sustain high Eddington ratios, MBH must be accreting gas that is both colder and denser than the typical inter-stellar medium, which must be sufficiently close to the MBH to be accreted onto the MBH accretion disc. In gas-rich disc galaxies, such gas is distributed throughout the disc in the form of giant molecular clouds. Observations have shown that extended molecular gas in the host galaxy does not correlate with AGN activity \citep{maiolino1997MolecularGasMorphology,saintonge2012IMPACTINTERACTIONSBARS}. Radiative AGN in the local Universe \citep{heckman2014CoevolutionGalaxiesSupermassive} and out to $z=5$ \citep{suh2019MultiwavelengthPropertiesType} are located in galaxies with an abundant cold dense gas supply, which is centrally concentrated \citep{hicks2013FUELINGACTIVEGALACTIC}. To grow a MBH, an evolutionary process which drives gas towards the central regions of a galaxy is therefore necessary. 

Analytic studies suggest that the outer edge of accretion discs around MBH are truncated due to self-gravity at scales of order 0.01 pc \citep{collin-souffrin1990LineContinuumEmission,shlosman1990FuellingActiveGalactic}. This means that to feed MBH, gas needs to travel from kpc scales in the host galaxy to sub-pc scales near the MBH. Such inflows do not happen in all environments. In the process, it needs to lose more than 99 \% of its angular momentum \citep{jogee2006FuelingEvolutionAGN}. As the existence of bulgeless galaxies shows, angular momentum can prevent inflows into galactic centres over very long periods of time. Large-scale gravitational torques are required to funnel gas from the galactic disc towards the centre, to grow pseudo-bulges, drive nuclear starbursts, and fuel AGN. While the long-held paradigm has been that galaxy mergers lead to the majority of MBH growth through a re-distribution of the angular momentum of gas in a galaxy, as discussed above, observational and theoretical studies in the past decade have revealed that internal, secular processes between mergers work in combination with accretion following mergers to build MBH mass over long periods of time. Below we discuss the different processes that have been proposed to drive MBH growth.

\subsection{The role of mergers}
\label{sec:merger_growth}
Simulations show that merging two gas-rich galaxies causes large-scale disc-instabilities that drive gas to the centre, fuelling a central starburst \citep{barnes1991FuelingStarburstGalaxies, barnes1996TransformationsGalaxiesII, mihos1996GasdynamicsStarburstsMajor}. Observationally, there is also strong evidence for links between star formation and mergers. ULIRGs (ultra-luminous infra-red galaxies) are massive dusty starbursts in the local Universe that are thought to be triggered by major mergers \citep{sanders1996LUMINOUSINFRAREDGALAXIES}. More broadly, \citet{li2008InteractioninducedStarFormation} studied the 2-point correlation function of local galaxies and report that major mergers compact galaxies and enhance star formation. In addition the redistribution of angular momentum in the merger leads to the formation of a dispersion supported bulge component in a galaxy \citep{tonini16, brooks2016BulgeFormationMergers}, even if a disk reforms post-merger \citep{SparreSpringel2017}. Mergers therefore clearly grow both total mass, bulge mass, and increase the velocity dispersion of a galaxy. 

Given the particularly tight correlations of MBH mass with galaxy bulge masses and bulge velocity dispersion, particularly for bulge-dominated galaxies (see Sec.~\ref{sec:scaling_relations}), this has been interpreted as evidence that major mergers drive coevolution  \citep{kormendy2013CoevolutionNotSupermassive}. It is not clear how much of the MBH mass in the local Universe has been accreted following major-merger driven accretion episodes.  Observationally, most ongoing MBH growth in the local Universe is not associated with major mergers \citep{darg2010GalaxyZooProperties,liu2011ActiveGalacticNucleus}. MBH in mergers do grow rapidly, and produce luminous AGN, but due to their rarity they supply only about 20\% of the total MBH mass growth in the local Universe \citep{wild2007BurstyStellarPopulations}. There is also evidence for a significant delay between the merger-induced starburst and the peak in AGN activity on the order of several hundred Myr \citep{schawinski2007ObservationalEvidenceAGN,davies2007CloseLookStar,wild2010TimingStarburstAGNConnection}. \citet{alexander2012WhatDrivesGrowth} suggest that this delay is evidence that post-starburst AGN are fuelled by the stellar winds from the star-burst formed stars but it could also simply be evidence that delays for cosmic gas to lose sufficient angular momentum to be accreted onto the MBH accretion disc are very long.

Semi-analytic models have had some success in reproducing general trends of the observed AGN luminosity function and MBH mass functions by assuming that MBH grow predominantly after major galaxy mergers \citep{kauffmann2000UnifiedModelEvolution,volonteri2003AssemblyMergingHistory,hopkins2008CosmologicalFrameworkCoevolution,malbon2007BlackHoleGrowth,bower2006BreakingHierarchyGalaxy}. However, such models struggle to produce massive BH as early as we observe them without over-producing massive BH at low redshift \citep{natarajan2012MassFunctionBlack}, and generally struggle to match the observed MBH mass function and luminosity function simultaneously \citep{ricarte2018ExploringSMBHAssembly,volonteri2008EvolutionMassiveBlack}. They also showed that the assumption that MBH grow at a constant efficiency compared to the star formation rate of the host galaxy fail to reproduce observed mass and luminosity functions \citep{haiman2004ReasoningFossilsLearning,granato2004PhysicalModelCoevolution,cattaneo2005ActiveGalacticNuclei}, whether using merger-driven star formation or star-formation more generally. 

The insights from simulations are more complex. Some of the first galaxy-scale simulations already showed that merging gas-rich spirals drive star bursts and MBH growth, and allow MBH to naturally grow onto the observed $\mbh -\ssigma$ relation \citep{dimatteo2005EnergyInputQuasars,capelo2017ShocksAngularMomentum}.  On the other hand, as discussed above, cosmological simulations show the onset of efficient MBH growth after the host galaxy reaches a threshold mass which does not require major mergers. \citet{sijacki2015IllustrisSimulationEvolving} conclude from their work using the Illustris simulation that the triggering of star formation and central AGN do not necessarily have a common origin or coherent timing, which contributes to the scatter in their simulated MBH-galaxy scaling relations. \citet{volonteri2022DualAGNHorizonAGN} report no significant increase in MBH accretion rate for an ongoing galaxy major merger, which is in agreement with observations that predict a significant time delay between merger-induced star burst and peak in AGN activity. \citet{ricarte2018ExploringSMBHAssembly} generally report no link between efficient MBH growth and major mergers. \citet{steinborn2018CosmologicalSimulationsBlack} show that the majority of rapid MBH growth phases follow a major merger but point out that this simply reflects the peak in merger rates for massive galaxies at the peak of MBH growth around $z=2$. By contrast, \citet{mcalpine2018RapidGrowthPhase} argue that up to 60 \% of their simulated rapid MBH growth phases are triggered by major mergers. Whereas \citet{mcalpine2020} conclude that although they find an increased abundance of luminous AGN within merging systems relative to control samples of isolated galaxies, galaxy mergers do not induce a significant amount of BH growth ($\sim15\%$).  

\subsection{Bars, spirals, and other non-axisymmetric features}
Large-scale non-axisymmetric perturbations in the stellar potential of the galaxy, which can take the form of bars, spirals and oval distortions \citep{shlosman1989EvolutionSelfGravitatingAccretion,sellwood2014SecularEvolutionDisk}, are also thought to funnel gas inwards. Bars can form secularly due to small perturbations in the disc \citep{athanassoula1986BisymmetricInstabilitiesKuzmin,ostriker1973NumericalStudyStability,bournaud2002GasAccretionSpiral}, or be triggered in otherwise stable discs by tidal interactions \citep{noguchi1996BarredGalaxiesIntrinsic}, or minor mergers \citep{bournaud2005GalaxyMergersVarious}. Such perturbations drive gas into the galactic centre \citep{shlosman1990FuellingActiveGalactic,yu2022EDGECALIFASurveyRole} at rates in excess of typical MBH accretion rates \citep{Sakamoto1996, Maciejewski2002, Regan2004, Lin2013, Smethurst2021}. Despite these strong predictions from simulations that barred galaxies host higher mass MBH with higher accretion rates \citep[e.g. in IllustrisTNG100][]{kataria2023HowDoesPresence}, there is an ongoing debate in the literature about the correlation of bars with AGN activity. 

Some studies of large-scale galaxy surveys have presented evidence that bars do not correlate with AGN activity \citep{martini2003CircumnuclearDustNearby,jiang2011BLACKHOLEMASS,lee2012BarsTriggerActivity, Cheung2015, Goudling2017}, finding that barred fractions in galaxies with and without AGN are approximately the same. In addition, AGN in barred galaxies frequently have very low Eddington ratios \citep[e.g][]{cisternas2013XRayNuclearActivity}. However, some studies do find a correlation between the presence of an AGN and the presence of a bar \citep{Knapen2000, Laine2002, Laurikainen2004, Coelho2011, Oh2012, Alonso2018, Garland2023}. For example, \citet{Galloway2015} and \citet{SilvaLima2022} both find a higher AGN fraction in barred galaxies compared to unbarred systems, but not higher AGN accretion rates.

There have been many explanations raised for this discrepancy between studies, some physical and some due to observational biases. Starting with the physical explanations, it has been suggested that bar-driven inflows do not reach far enough into the centre of a galaxy to feed the AGN \citep{garcia-burillo2005MolecularGasNUclei}. Using simulations, \citet{fanali2015BarFormationDriver} showed that bars drive the largest inflows during formation, but that gas stalls far from the BH once the bar is established. A model of "bars-within-bars" was proposed to drive gas-flows further into the centre \citep{englmaier2004DynamicalDecouplingNested}. Using a series of nested zoom simulations, \citet{hopkins2010HowMassiveBlack} report that MBH are fed by a variety of structures within the galaxy, including spirals, rings, clumps, bars within bars, and nuclear spirals; each of which proved short-lived. This would explain why simulations like IllustrisTNG simulation report that barred galaxies have on average higher MBH masses \citep{kataria2023HowDoesPresence}, and the disagreement in the literature if secular feeding has short duty cycles.

However, it also been suggested that observational biases are responsible for the apparent disagreement in the literature. The combination of the rarity of AGN, the challenges separating the AGN emission from that of the host galaxy, the dependency on mass, colour, and bulge strength of barred galaxies, and the rarity of isolated disks which are free of the influences of mergers (e.g. bulge formation) may lead to a weakening of any apparent connection between MBH and bars. Bar strength is also thought to be an important factor, with \citet{Garland2024} showing how strongly barred galaxies are more likely to host an AGN than weakly barred galaxies, which are in turn more likely to host an AGN than unbarred systems (a $>5\sigma$ result); again suggesting that if studies do not control for the strength of the bar, then this will likely weaken any correlation.  Bar strength is also thought to be tied to the bar kinematics, i.e. whether a bar is a fast bar with the ends of the bar rotating at the same speed as the stars at that radius, or a slow bar where the ends rotate at a slower speed. \citet{geron2024} showed how the slowest, strongest bars have been shown to affect their host galaxy's star formation the most, suggesting that they cause the biggest impact on the gas in a galaxy. Further work is needed to establish the connection between the kinematics of bars and AGN hosting, however this field is limited by the sample sizes of bars with kinematic measurements.  

Simulations and observations have found that spiral arms are also able to funnel gas to the centres of galaxies in excess of observed MBH accretion rates \citep{maciejewski2004NuclearSpiralsGalaxies, Davies2009, SchnorrMuller2014, Slater2019, Smethurst2021}. In addition, a scaling relation (albeit one with a large scatter) is found between the pitch angle of spiral arms (i.e. the tightness of the spiral winding) and the mass of the MBH in spiral galaxies \citep{Seigar2008, Berrier2013, Davis2017}. The mass of the MBH increases as the spiral arm pitch angle decreases, or as the spirals become more tightly wound. Given that spiral winding is also correlated with bulge strength on the Hubble Sequence \citep{Hubble1926}, this scaling relation is not unexpected given the correlation between MBH mass and bulge mass discussed above. Therefore, as with studies which study the connection between bars and AGN, one must also control for bulge mass when considering the spiral-AGN connection. 

%However, more recent observational studies controlling for parameters such as mass, colour, and bulge mass have found an excess of AGN in barred galaxies \citep{cite}. \rjs{Will discuss strong vs weak bar here (e.g. Izzy work), plus fast vs slow bars and kinematics of bars. Will flesh out this whole section eventually to have more support for bar feeding from observations and simulations. Also need to add spiral arm arguments as well, e.g. the fact large scatter pitch angle scaling relation exists}. \rbc{Sounds good, I will leave this to you}
%DONE 2-12-24

\section{The impact of black hole growth on their host galaxies}
\label{sec:impact_on_galaxies}

\begin{figure}
    \centering
    \includegraphics[width=\linewidth]{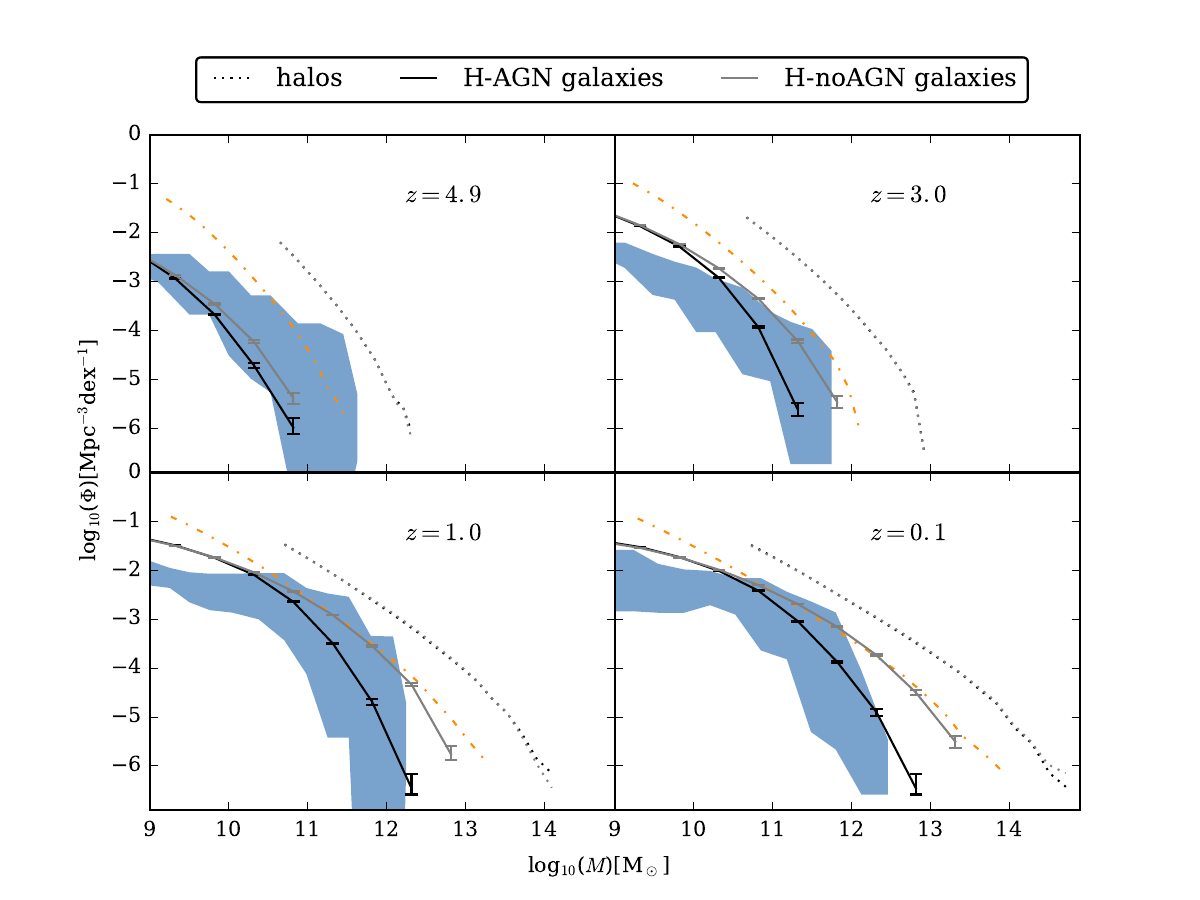}
    \caption{Galaxy stellar mass function from the Horizon-AGN (black, with AGN feedback) and Horizon-noAGN (grey, without AGN feedback) simulations. Simulated data is compared to combined observations (blue shaded) and the expected distribution from the dark matter halo mass function assuming a universal baryon fraction of 16.5 \% (orange dashed). This plot shows that at low redshift the observed galaxy stellar mass functions (blue) have a significantly steeper slope than would be expected from the dark matter halo mass function (orange dashed). It is only the simulation with feedback that is able to match the observations at the high-mass end. Reproduced from \citet{beckmann2017CosmicEvolutionStellar}. }
    \label{fig:GSMF}
\end{figure}

One of the core challenges of the current paradigm of a $\Lambda$-CDM cosmology is that the mass function of massive dark matter halos and galaxies have different slopes \citep{white1991GalaxyFormationHierarchical,moster2010ConstraintsRelationshipStellar} (see Fig.~\ref{fig:GSMF}). This means that the ratio of galaxy stellar mass to dark matter halo mass ($\mstar / M_{\rm DM}$) is lower for high-mass galaxies which in turn means that massive galaxies must be less efficient at forming stars at late times \citep{mutch2013SimplestModelGalaxy,behroozi2013AverageStarFormation}. The lack of star formation in massive galaxies \citep{mcdonald2018RevisitingCoolingFlow} is particularly puzzling. In massive galaxy cluster environments the hot intracluster medium of large galaxy clusters emits strongly in the X-ray, suggesting high cooling rates. This should produce a mass inflow onto the central cluster of hundreds to thousands of solar masses per year which would fuel massive starbursts \citep{kauffmann2000UnifiedModelEvolution}. Instead, bulge-dominated central cluster galaxies show little sign of ongoing star-formation. This is known as the "cooling flow problem" \citep[see][for a review]{fabian1994CoolingFlowsClusters}. The observational evidence suggests that there must be a mechanism that reduces star formation once galaxies grow above a threshold mass, and then keeps it low over long periods of time.

Early semi-analytic models of galaxy evolution already showed that supernovae provide insufficient energy to suppress the massive end of the galaxy stellar mass function in line with observations, especially if stars are allowed to form in galaxy cluster cooling flows \citep{kauffmann2000UnifiedModelEvolution}. AGN feedback provides a natural second source of energy injection. Early semi-analytic models found good agreement with observations by assuming that the cold gas supply in galaxies is destroyed by the supernova feedback from galaxy-merger driven starbusts (which also efficiently grow MBH), and that star formation is suppressed in the long run by AGN feedback that offsets the energy lost in cooling flows \citep{croton2006ManyLivesActive,bower2006BreakingHierarchyGalaxy,somerville2008SemianalyticModelCoevolution}. In this model, AGN predominantly provide so-called ``maintenance-mode'' feedback, whereby they keep already quenched galaxies red and dead by offsetting any long-term cooling losses. 

The ``AGN maintenance mode'' hypothesis is supported by nearly universal radio emission observed at the centre of local bulge-dominated galaxies \citep{burns1990RadioPropertiesCD,sabater2019LoTSSViewRadio}. About 50\% of sources are central only, while the remaining sources show signs of  extended jets or diffuse radio sources \citep{kolokythas2018CompleteLocalvolumeGroups}. This radio emission is thought to be the observational signature of near-universal AGN-driven jets launched by AGN in a radiatively inefficient accretion mode (See Sec.~\ref{sec:accretion_states} and Sec.~\ref{sec:MBHgrowthlocal}), which transfer energy to the intracluster medium by inflating the radio bubbles that can be observed in nearby galaxy groups and clusters with short cooling times \citep{dunn2006InvestigatingAGNHeating, hlavacek-larrondo2022AGNFeedbackGroups}. The mechanical power of the bubbles matches the observed cooling rates \citep{best2006AGNcontrolledCoolingElliptical}, or even exceeds them for low-mass systems \citep{nulsen2007AGNHeatingCavities}, providing further support for AGN maintenance feedback \citep{fabian2012ObservationalEvidenceActive}. Simulations of galaxy clusters have shown that AGN jets can efficiently regulate cooling flows in massive galaxy clusters over long periods \citep{yang2016HowAGNJets, prasad2018CoolcoreClustersRole, beckmann2019DenseGasFormation}, especially if supported by other heating mechanisms such as turbulence \citep{kunz2011ThermallyStableHeating,voit2018RoleTurbulenceCircumgalactic} and cosmic rays \citep{sijacki2008SimulationsCosmicrayFeedback,ehlert2018SimulationsDynamicsMagnetized,beckmann2022CosmicRaysThermal}. Such AGN are fed by series of chaotic cold clouds \citep{king2007FuellingActiveGalactic, gaspari2013ChaoticColdAccretion}  that condenses from the hot intracluster through local thermal instability \citep{mccourt2012ThermalInstabilityGravitationally}. When cold gas forms the central AGN is fed and drives powerful jets that inflate hot bubbles which heat the intracluster-medium. This energy injection prevents global thermal instability (i.e. prevents the cooling flow) and destroys or uplifts the central cold clouds which cuts off the AGN fuel supply and turns off the AGN jet until the next cycle. In this way, the large-scale thermal evolution of the cluster and the central AGN are balanced in a self-regulating feedback where more cooling leads to more AGN activity and vice-versa. 

Cosmological simulations resolve the three-dimensional structure of cold gas in galaxies, unlike in semi-analytic models,  and quickly showed that the impact of star formation is insufficient to quench star formation. Instead it has been hypothesised that efficiently accreting ``quasar'' mode AGN might drive galaxy-scale winds that quench star formation. Observationally, AGN host galaxies often show evidence of multi-phase galactic winds from scales of the central hundreds of pc \citep{garcia-burillo2021GalaxyActivityTorus} to kpc scales \citep{genzel2014EvidenceWidespreadActive,woo2016PrevalenceGasOutflows,rupke2019100kiloparsecWindFeeding}. The fraction of AGN with a signature of outflow kinematics, steeply increases with AGN luminosity and Eddington ratio \citep{woo2017DelayedNoFeedback}. The sizes and inferred outflow properties of such winds seem to scale with AGN luminosity $L_{\rm AGN}$ \citep{veilleux2013FastMolecularOutflows,cicone2014MassiveMolecularOutflows,fiore2017AGNWindScaling,wylezalek2020IonizedGasOutflow}, as far as can be determined given the difficulties in observationally constraining the spatial extent, outflow velocity and the outflow mass of such galaxy-scale winds \citep{harrison2018AGNOutflowsFeedback}. As a result of these difficulties, the error bars on the outflow power can span several orders of magnitude, particularly for multi-phase outflows \citep{cicone2018LargelyUnconstrainedMultiphase, villar-martin2016IonizedOutflowsLuminous, Smethurst2021}. It is much more difficult to determine if and how such winds affect star formation. If galactic winds efficiently and quickly quenched star formation then we would expect galaxies with the strongest AGN (and winds) to show the least amount of star formation. As discussed in Section~\ref{sec:MBHgrowthlocal}, in the local Universe the opposite appears to be the case: higher AGN luminosities are generally found in galaxies with higher star formation rates, and the strongest bursts of AGN activity following major galaxy mergers lag behind the merger-drive star-bursts, not vice versa. 

By contrast, in cosmological simulations that allow for AGN feedback energy to couple to the interstellar medium at all $\fedd$, using either a two-mode feedback mode with different coupling for ``quasar mode'' and ``jet mode'' \citep{sijacki2007UnifiedModelAGN,sijacki2015IllustrisSimulationEvolving,weinberger2017SimulatingGalaxyFormation,dubois2014DancingDarkGalactic} or just a single coupling mode applicable at all $\fedd$ \citep{schaye2015EAGLEProjectSimulating,mcalpine2018RapidGrowthPhase,dave2019SIMBACosmologicalSimulations}. Such AGN feedback has been shown to be crucial to reproduce a whole range of properties of the observed population of galaxies including matching the high-mass end of the galaxy stellar mass function (see also Fig.~\ref{fig:GSMF}) \citep{dubois2014DancingDarkGalactic,schaye2015EAGLEProjectSimulating,pillepich2018SimulatingGalaxyFormation}, the observed colour-bimodiality of galaxies \citep{kaviraj2017HorizonAGNSimulationEvolution,trayford2015ColoursLuminosities01,nelson2018FirstResultsIllustrisTNG}, the kinematic structure and size evolution of massive galaxies \citep{dubois2013AGNdrivenQuenchingStar,peirani2017DensityProfileDark,choi2018RoleBlackHole}, chemical abundance patterns of stellar populations \citep{taylor2015EffectsAGNFeedback,segers2016OriginAenhancementMassive} and the self-regulation of MBH growth \citep{dimatteo2005EnergyInputQuasars,angles-alcazar2015TORQUELIMITEDGROWTHMASSIVE,volonteri2016CosmicEvolutionMassive}. In simulations AGN feedback acts by driving the observed galaxy-scale winds but it also influences galaxies by reducing inflows into massive galaxies \citep{beckmann2017CosmicEvolutionStellar} which controls long-term star-formation \citep{dave2020GalaxyColdGas}.

The model of AGN-feedback regulated galaxy evolution is not perfect. Apart from the difficulties of establishing clear observational evidence for AGN quenching of star formation \citep{smethurst2016}, and remaining discrepancies between simulated and observed MBH and galaxy populations, there are also concerns on the theoretical side. One difficulty with our current model of AGN feedback in cosmological simulations is that it is very unclear how the energy couples to the interstellar medium \citep{costa2020PoweringGalacticSuperwinds,vogelsberger2020CosmologicalSimulationsGalaxy} and also that simulations predict very different distributions of star formation rates as a function of MBH and galaxy stellar mass \citep{habouzit2021SupermassiveBlackHoles} which means that they disagree on when and where AGN feedback quenches star formation. 

\section{Summary of our model of coevolution}
In this chapter we presented evidence why MBH and galaxies are thought to tightly coevolve. 

\begin{enumerate}
\item \textbf{Prevalence of MBH} Almost every galaxy hosts a massive black hole (MBH), whose properties tightly correlating with those of their host galaxy. 
\item \textbf{Correlations between MBH and galaxy properties} MBH and galaxies demonstrate strong scaling relations, such as those between MBH mass and galaxy stellar velocity dispersion. These correlations suggest a coevolution  between MBH and galaxies. There are variations in the scaling relations with AGN and galaxy properties which informs us about distinct evolutionary paths.
\item \textbf{Coevolution across cosmic time} The coevolution spans most of cosmic history. Observations indicate that the first active MBH, seen as active galactic nuclei (AGN), emerged as early as redshift $z \sim 10$. This early emergence indicates that the mechanisms driving coevolution have been at play for a substantial part of the universe's history.
\item \textbf{MBH growth across cosmic time} MBH growth is determined by the central cold gas supply of the host galaxy. Large-scale galactic torques are needed to drive cold gas towards the MBH. MBH is periodic, which active periods interspersed with times of quiescence. MBH undergo cosmic downsizing: Since cosmic noon, around $z\sim 2$, average accretion rates have fallen and the mass of the most active MBH has decreased. As a result, the most rapidly growing AGN in the local Universe are powered by MBH in the mass range $\mbh = 10^{6-8} \rm \ \Msun $.
\item \textbf{The two channels of coevolution} MBH accretion efficiency is quantified using the Eddington ratio $\fedd$. At $\fedd > 0.01$, MBH are fed by a geometrically thick, optically thin accretion disc that emits feedback energy predominently in the form of radiation. At $\fedd < 0.01$, MBH accrete via a geometrically thick, optically thin accretion disc that drives powerful jets. 
\item \textbf{What drives MBH growth}  MBH are fed through a combination of galaxy-internal, secular processes and those driven by interactions between galaxies. There is no compelling evidence that either bars or galaxy mergers dominate MBH evolution and drive coevolution.
\item \textbf{AGN feedback regulates MBH growth and star formation} When growing, MBH convert some of the accreted mass into feedback energy, which is reinjected into the surrounding of the galaxy and couples to the gas supply in the host galaxy across multiple scales. In this way, AGN regulate their own mass growth and star formation in their host galaxy. As they and their host galaxies become more massive, MBH transition from radiatively efficient accretion to radiatively-inefficient accretion that drives jet-based maintenance mode feedback.
\item \textbf{Understanding coevolution} Cosmological simulations provide the most powerful test of our model of MBH and galaxy coevolution. Current simulations reproduce many features of the observed populations of MBH and galaxies. Remaining discrepancies between simulated and observed MBH and galaxies offer an opportunity to expand our model of MBH growth and MBH-galaxy coevolution.
\end{enumerate}

While our current understanding of M BH growth, and the AGN-feedback based model of MBH-galaxy coevolution are able to reproduce many features of the observed population of MBH and galaxies, plentiful open questions remain. Particular areas of research at the moment include the formation of seed black holes, the growth of MBH in the early Universe and in dwarf galaxies, the unknown distribution of MBH spin and its impact on AGN feedback efficiencies across cosmic time, and the details of MBH dynamics in and around host galaxies. 

\section*{Acknowledgements}
R.S. Beckmann acknowledges funding from UKRI Future Leaders Fellowship, grant number MR/Y015517/1. R.J. Smethurst gratefully acknowledges funding from the Royal Astronomical Society. This is a pre-print of a chapter for the Encyclopedia of Astrophysics (edited by I. Mandel, section editor S. McGee) to be published by Elsevier as a Reference Module. For the purpose of open access, the author has applied a Creative Commons Attribution (CC BY) licence to any Author Accepted Manuscript version arising from this submission. 

%\printbibliography

\bibliographystyle{Harvard}
%\bibliography{ricarda_paper_library,papers_becky}
\bibliography{paper_library}

\end{document}